\documentclass[prx,aps,twocolumn,superscriptaddress,longbibliography]{revtex4-1}
\usepackage{graphicx}
\usepackage{amsmath}
\usepackage{amssymb}
\usepackage{color}
\usepackage{comment}

\newcommand{\beginsupplement}{%
        \setcounter{table}{0}
        \renewcommand{\thetable}{S\arabic{table}}%
        \setcounter{figure}{0}
        \renewcommand{\thefigure}{S\arabic{figure}}%
     }

\begin{document}

\title{Water-like anomalies as a function of tetrahedrality}
\author{John Russo}
\affiliation{School of Mathematics, University of Bristol, Bristol BS8 1TW, UK }
\affiliation{Institute of Industrial Science, University of Tokyo, 4-6-1 Komaba, Meguro-ku, Tokyo 153-8505, Japan}
\author{Kenji Akahane}
\affiliation{Institute of Industrial Science, University of Tokyo, 4-6-1 Komaba, Meguro-ku, Tokyo 153-8505, Japan}
\author{Hajime Tanaka}
\email{Corresponding author: tanaka@iis.u-tokyo.ac.jp}
\affiliation{Institute of Industrial Science, University of Tokyo, 4-6-1 Komaba, Meguro-ku, Tokyo 153-8505, Japan}

\date{\today}

\begin{abstract}
Tetrahedral interactions describe the behaviour of the most abundant and technologically important materials on Earth, such as water, silicon, carbon, germanium, and countless others. Despite their differences, these materials share unique common physical behaviours, such as liquid anomalies, open crystalline structures, and extremely poor glass-forming ability at ambient pressure. To reveal the physical origin of these anomalies and their link to the shape of the phase diagram, we systematically study the properties of the Stillinger-Weber potential as a function of the strength of the tetrahedral interaction $\lambda$.
We uncover a new transition to a re-entrant spinodal line at low values of $\lambda$, accompanied with a change in the dynamical behaviour, from Non-Arrhenius to Arrhenius.
We then show that a two-state model can provide a comprehensive understanding on how the thermodynamic and dynamic anomalies of this important class of materials depend on the strength of the tetrahedral interaction. Our work establishes a deep link between the shape of phase diagram and the thermodynamic and dynamic properties through local structural ordering in liquids, and hints at why water is so special among all substances.
\end{abstract}

\date{\today}

\maketitle

Liquids do not possess long-range order but often have short-range order. For example, water, silicon, germanium, and carbon are known to form tetrahedral order locally because of the directional nature of hydrogen or covalent bonding. These liquids commonly exhibit anomalous thermodynamic and dynamic anomalies, which are absent in ordinary liquids, e.g., van-der-Waals liquids. Interestingly, all these tetrahedral liquids also have unusual V-shaped phase diagrams.
Liquid anomalies include the density maximum as a function of temperature $T$, the steep increase in the isothermal compressibility and heat capacity 
upon cooling, the non-Arrhenius behaviour of viscosity and diffusion constant at low pressures, and 
the minimum of viscosity and the maximum of diffusion constant as a function of pressure $P$ (see, e.g., Refs.~ \cite{debenedetti1996metastable,Mishima1998,debenedetti2003supercooled,Nilsson2015,gallo2016water} for water anomalies).  
Furthermore, all these liquids commonly have V-shaped $P$-$T$ solid-liquid phase diagrams, 
in which the melting point has a minimum at 
a positive pressure $P_x$. It was argued \cite{TanakaWPRB} that there is a deep link between the shape of the phase diagram and these anomalous thermodynamic and kinetic behaviours, as a consequence of local tetrahedral ordering. 
However, it has remained elusive how the degree of tetrahedrality controls the shape of phase diagram 
and the anomalies. To address this problem, we need a model where we can control tetrahedrality in a systematic manner.

As a coarse-grained classical model for tetrahedral materials, the Stillinger-Weber (SW) potential has emerged
as an effective model potential capable of capturing all the relevant physical properties that stem
from the tetrahedrality of the interactions. 
The original parameterization
of the SW potential was targeted to the bulk properties of silicon~\cite{stillingerweber},
but has also found widespread applicability in the modeling of other Group XIV elements~\cite{molinero2008water}.
Apart from atomic fluids, the SW potential has also found application in the coarse-grained
description of complex molecular fluids. The most notable example is water,
whose SW representation is intermediate between that of
silicon and carbon, and is known as mW water~\cite{molinero2008water}.
While retaining a high degree of structural accuracy, the mW model has proven
to be very efficient from a computational point of view, and has played
a big role in the study of water crystallization~\cite{molinero_nature,reinhardt2012free,li2013ice,russo2014new,sosso2016crystal}, which otherwise requires advanced techniques~\cite{sosso2016microscopic,pipolo2017navigating}.
As a good model of water, the mW water exhibits a vast array of thermodynamic and dynamic anomalies~\cite{limmer_molinero,sengupta2014diffusivity,singh2014triplet,dhabal2015excess},
and recently the behaviour of the anomalies for different values of $\lambda$ was considered in Refs.~\cite{angell2016potential,dhabal}.
Ref.~\cite{angell2016potential} focused on the location of the second critical point, studied by means of the \emph{isochore crossing} technique, showing that changing $\lambda$ can decrease the critical pressure to ambient conditions, and down to the liquid-vapour spinodal, as in the \emph{critical point-free scenario}. In Ref.~\cite{dhabal}, the full hierarchy of anomalies was considered for three different values of $\lambda$, showing that they follow a silica-like hierarchy, which becomes a water-like hierarchy if the excess entropy and Rosenfeld scaling are considered.
These works have shown the richness of the behaviour of the SW model, and opened the question on whether we can rationalize
the anomalous behaviour of tetrahedral liquids, and if we can connect their behaviour to the underlying phase diagram.

In this Article we consider the anomalous behaviour of the liquid phase, focusing in particular on the liquid anomalies
that occur at negative pressures. Our goal is to connect the behaviour of liquid anomalies with the change of thermodynamic properties as a function of $\lambda$. We start by computing the full phase diagrams in the extended ($T$, $P$, $\lambda$) thermodynamic space, extending the results of Ref.~\cite{akahane_quadruple} to negative pressures, where clathrate structures are the stable crystals. The region at negative pressure is crucial for unveiling the origin of the anomalous behaviour, and by measuring the density fluctuations,
we track the stability limit of the liquid, i.e. the liquid-to-gas spinodal line. We find evidence for a transition from a positively-sloped spinodal line to a re-entrant spinodal as a function of the $\lambda$ parameter, providing the first example of such a transition in a water-like model. We show how this result is connected to the anomalous phase behaviour of water, and argue that a two-state modeling of the liquid phase \cite{tanaka1998simple,tanaka2000simple,Tanaka2000a,Tanaka2003,TanakaWPRB,russo2014understanding,holten2012entropy,
limmer_molinero,holten2013nature,holten2014two} provides a simple theoretical framework which rationalizes the anomalous behaviour of tetrahedral liquids as a function of the strength of the tetrahedral interaction. 
The model provides a deep link between the anomalies and the shape of the phase diagram, as a consequence of the fact that locally favoured structures in liquids have the same local symmetry as the low-pressure diamond crystal. More precisely, we reveal that the value of $\lambda$ corresponding to water maximizes two-state features and the resulting anomalies, providing structural flexibility to water: 
water can change its physical and chemical properties by changing an extra structural degree of freedom, i.e., 
the fraction of the two states, in response to external perturbations. 

\section*{Results}
\subsection*{Two-state model}\label{sec:twostate}

Liquid anomalies can be divided in two categories: thermodynamic and dynamic anomalies. Thermodynamic ones originate from the anomalous temperature dependence of a thermodynamic response function: unlike the ordinary behaviour of simple liquids, in water, thermodynamic fluctuations show an increase with lowering the temperature. An example is given by the isothermal compressibility $\kappa_T$, which is proportional to volume fluctuations, and displays a minimum at around $T=319$~K, below which it shows a rapid increase. Similar anomalies are shown by the density $\rho$ (which has a maximum at $T=277.15 K$), and the specific heat $C_p$ (which has a minimum around $T=308\,K$). 

In order to rationalize both thermodynamic and dynamic anomalies, we employ a two-state model. The history of two-state models of water dates back to R{\"o}ntgen~\cite{rontgen1892ueber}. The basic idea is that the anomalies of water can be understood if water is described as a mixture of two components in thermodynamic equilibrium, such that the concentration of the mixture is state-dependent. Until recently, however, water was described as a mixture of distinct structural components, whose number is two \cite{Angell_w,robinson,ponyatovsky} to four  \cite{nemethy1962structure}.
 
Only recently, the importance of the degeneracy of states (or, the large entropic loss upon the formation of locally favoured tetrahedral structures) 
was properly recognized \cite{tanaka1998simple,tanaka2000simple,Tanaka2000a}. Ref.~\cite{anisimov2018thermodynamics} discusses the connection between two-state models and fluid polyamorphism in a general way in a variety of condensed matter systems.
Furthermore, unlike previous approaches, where the order parameter is only density, it was proposed \cite{tanaka_review,tanaka1999two} that we need at least two order parameters to understand the phenomena: 
one is the density $\rho$ and the other is bond order $s$, which represents the local break-down of rotational symmetry due to directional bonding. This bond order is also associated with the rotational symmetry that is broken upon crystallization, which is the key to a link between the two-state behaviour and the phase diagram. 
The order parameter $s$ is defined as the fraction of locally favoured structures. The importance of the two-order-parameter description was verified for model water by numerical simulations \cite{errington2001relationship}. 
Note that the density order parameter is conserved, but the bond order parameter is not since 
locally favoured structures can be created and annihilated locally. 
This idea has been formalized by writing the free energy of water as that of a regular mixture of two components with very different degeneracy of states, under the additional equilibrium condition between the two components. This has produced a family of models that are often employed to fit water's equation of state with high precision~\cite{tanaka1998simple,tanaka2000simple,Tanaka2000a,Tanaka2003,TanakaWPRB,holten2012entropy,russo2014understanding,limmer_molinero,holten2014two}. Recent approaches go beyond the phenomenological use of two-state equation of states, and attempt to derive a two-state description starting from microscopic structural information~\cite{cuthbertson2011mixturelike,wikfeldt2011spatially,holten2013nature}. In our approach~\cite{russo2014understanding}, we identify the two states according to the degree of translational order up to the second shell. By introducing a structural parameter that measures translational order (that we call $\zeta$), we divide the population of water molecules into two collections of states: the $S$-state, comprising highly ordered states, where there is a clear separation between first and second shell of nearest neighbours, and the $\rho$-state, low ordered states characterized by a disordered arrangements of second shell molecules, including configurations with shell interpenetration. We define $s$ as the fraction of $S$ states, which at any given $T$ and $P$ can be written, provided that there is little cooperativity in formation of locally favoured structures, as \cite{tanaka2000simple,Tanaka2000a}
\begin{equation}\label{eq:s}
 s=\frac{g\exp{\beta(\Delta E-P\Delta v)}}{1+g\exp{\beta(\Delta E-P\Delta v)}}, 
\end{equation}
where $\Delta E=E_\rho-E_S$ is the energy difference between the $S$ and $\rho$ states, $\Delta v=v_S-v_\rho$ is their specific volume difference, and $g$ is a measure of the degeneracy of the $S$ state compared to the degeneracy of the $\rho$ state ($\Delta \sigma=k_{\rm B}\ln g$, where $\Delta \sigma$ is the entropy difference between the two states). The fraction of $S$ states controls the degree of anomalous behaviour of the mixture. Following the notation of Ref.~\cite{tanaka2012bond}, the specific volume is then given by
\begin{equation}\label{eq:v}
 v(T,P)=a(P)T+b(P)+s\Delta v,
\end{equation}
and the isothermal compressibility by
\begin{equation}\label{eq:k}
\kappa_T(T,P)=k(P)T^2+n(P)+sC(P),
\end{equation}
where the first two terms in each equation ($a(P)$, $b(P)$, and $k(P)$, $n(P)$), represent the background behaviour, and are obtained by fitting the specific volume and the compressibility far from the anomalous region. 
In the framework of the two-state model, the Widom-line is nothing but the equimolar line $s=1/2$,  or the the line of the Schottky anomaly~\cite{tanaka2012bond}, and can be written as
\begin{equation}\label{eq:widom}
 T_\text{W}=-\frac{\Delta E-P\Delta v}{\ln g}.
\end{equation}

Note that two-state model predictions can accommodate a liquid-liquid critical point through a positive free enthalpy of mixing term, but the Schottky anomalies arise whether or not this term is present or not.
In this work we set the enthalpy of mixing to zero ($J=0$ in the notation of Ref.~\cite{tanaka2012bond,holten2012entropy}), as it produces the best results, also in line with what observed for the mW model in Ref.~\cite{limmer_molinero}.

One compelling feature of our two-state model is that it can describe thermodynamic and dynamic anomalies ~\cite{tanaka2000simple,Tanaka2000a,Tanaka2003}. In the case of dynamical anomalies, the predictions of two-state models are remarkably different from alternative explanations of dynamic anomalies. The major contender to the description of dynamic anomalies is based on glassy phenomenology, which is known as the \emph{fragile-to-strong} transition~\cite{xu2005relation,cerveny2016confined,faraone2004fragile,zhang2009dynamic,gallo2010dynamic,wang2015dynamic}.

In the case of our two-state model~\cite{tanaka2000simple,Tanaka2000a,Tanaka2003}, instead, the two different states have different activation energies, $E^a_\rho$ and $E^a_S$ (with $\Delta E^a=E^a_S-E^a_\rho$), and the diffusion process can be written as
\begin{equation}\label{eq:D}
D=D_0\exp\left[-\frac{E^a_\rho+\bar{s}\Delta E^a}{k_{\rm B}T}\right], 
\end{equation}
where $\bar{s}$ is the fraction of dynamic $S$ state. It was assumed~\cite{tanaka2012bond} that $\bar s=s$, i.e. that the dynamic and static fractions of the $S$ states coincide. We will adopt this assumption here. However, we note that for an accurate description of dynamic anomalies a \emph{hierarchical} two-state model has to be considered~\cite{shi2017}. In the case of $g\ll 1$, corresponding to a much lower degeneracy of the $S$ state compared to the $\rho$ state, the expression (\ref{eq:s}) can be approximated as~\cite{tanaka2000simple,Tanaka2000a,Tanaka2003}
\begin{equation}
 s=g\exp\beta(\Delta E-P\Delta v). 
\end{equation}
Substituting this expression in Eq.~(\ref{eq:D}), and expanding to second order in $\beta$ (high $T$ expansion), we get
\begin{eqnarray}\label{eq:expansion}
 \ln(1/D)&\sim &\ln(1/D_0)+\beta\left[E^a_\rho+g\Delta E^a\right]+ \nonumber \\
 && +\beta^2\left[\Delta E^a g(\Delta E-P\Delta v)\right]. 
\end{eqnarray}

Equation~(\ref{eq:D}) predicts a full \emph{strong-to-strong} transition, from activation energy $E^a_\rho$ to $E^a_S$, 
instead of a fragile-to-strong transition. For small values of $s$, the crossover between the two strong behaviours can be fitted quadratically, with a coefficient that is proportional to $\Delta E^a$, i.e. the difference in the activation energy between the two states, $S$ and $\rho$.

In Ref.~\cite{singh2017pressure} a two-state model has been proposed in which the $\rho$ state behaves like a fragile liquid. Here we take a different approach, and provide evidence that a pure $\rho$-state behaves as a strong liquid in our model.

\subsection*{Generalised SW model}

The SW potential is composed of the sum of a pairwise term $U_2$ and three-body interaction term $U_3$ (see the Methods section for the definition of these terms): 
\begin{equation}\label{eq:sw}
U=\sum_{i}\sum_{j>i}U_2({\bf r}_{ij})+\lambda\sum_{i}\sum_{j\neq i}\sum_{k>j}U_3({\bf r}_{ij},{\bf r}_{jk}).
\end{equation}

Therefore, $\lambda$ is only parameter which differentiates the models.
$\lambda$ is a dimensionless parameter controlling the relative strength between pairwise and three-body term. For the mW model of water~\cite{molinero2008water} the value is $\lambda=23.15$, while for silicon the original parameter is $\lambda=21.0$~\cite{stillingerweber}. 
Furthermore, germanium and carbon are described by $\lambda$=20.0 and 26.2 respectively.

\begin{figure}[!t]
 \centering
 \includegraphics[width=7cm]{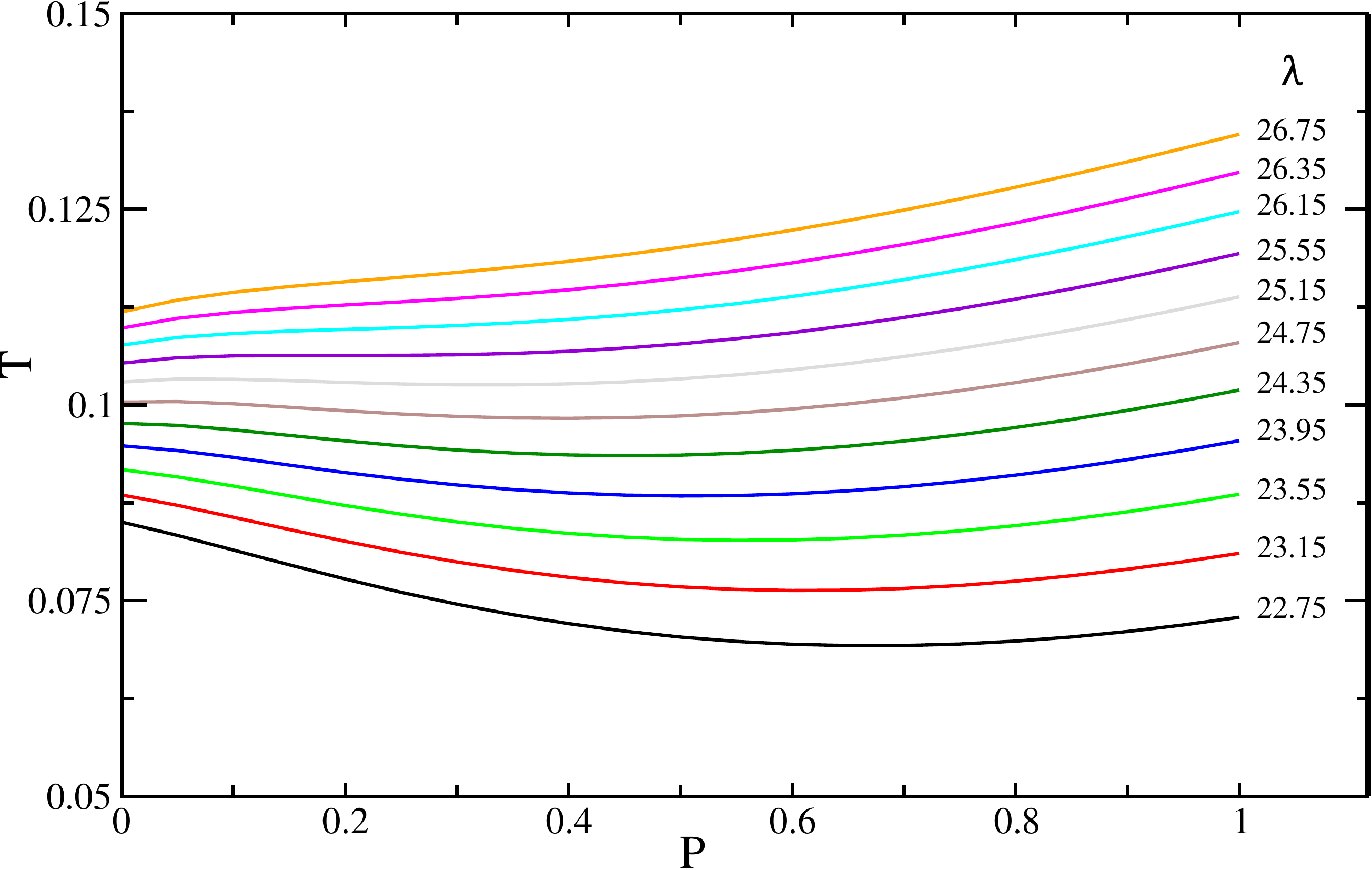}
 \caption{{\bf Melting lines of the $dc$ crystal at different values of $\lambda$.} With increasing $\lambda$ the temperature of the melting line increases, and the slope at $P=0$ goes from negative to positive, i.e. the crystal becomes more dense that the liquid.}
 \label{fig:melting}
\end{figure}
By tuning $\lambda$ one can continuously interpolate between the behaviour of water-like materials and the behaviour of simple fluids. To demonstrate this, in Fig.~\ref{fig:melting} we plot the melting line of the stable crystalline phase (the diamond cubic, dc, crystal) as a function of $\lambda$. The dc crystal is the only phase with a negatively sloped coexistence line, and the slope decreases with increasing $\lambda$, signaling the increase of density of the dc phase.
Eventually at high $\lambda$ the slope at $P=0$ becomes positive, when the diamond phase becomes more dense than the liquid. Figure ~\ref{fig:melting} shows that the change of slope occurs around $\lambda\sim 25$.
Thus, the V-shape feature of the phase diagram with $\partial T_{\rm m}/\partial P|_{P=0}<0$ exists only in a limited range of $\lambda$, i.e. $16<\lambda<25$. As shown later, this range roughly correspond to the region where we see water-like anomalies.

In Supplementary Information we plot the full phase diagram of the model, extending the results of Ref.~\cite{akahane_quadruple} to negative pressures.
Negative pressures are of great interest for at least two important reasons: 1) they stabilize clathrate lattices, which are crystalline structures with voids that can accommodate guest molecules, and are studied for energy storage, carbon dioxide sequestration, separation and natural gas storage~\cite{florusse2004stable,lee2005tuning,chatti2005benefits,struzhkin2007hydrogen}; 2) contrasting theories of the thermodynamic anomalies (in particular for the case of water) can be tested in the negative pressure region, both numerically and experimentally~\cite{azouzi2013coherent,pallares2014anomalies,gonzalez2016comprehensive,holten2017compressibility}.
In Supplementary Information we show that, at negative pressure, the BCC (body-centered cubic) phase is stable at lower $\lambda$ and the Si34 phase is stable at higher $\lambda$. In the following sections we will focus extensively on the line of liquid stability at negative pressures (the so-called spinodal). As a preliminary calculation, in Supplementary Information we have mapped the location of the critical point (from which the spinodal emanates) for a large range of values of $\lambda$ and reveal that, increasing the tetrahedral parameter $\lambda$ results in a lowering of both the critical temperature and pressure. As we will see later, this gives rise to a retracing spinodal~\cite{speedy1982stability} at low values of $\lambda$, when the spinodal line meets the line of density maxima.


\subsection*{Thermodynamic anomalies}

\begin{figure*}[!ht]
 \centering
 \includegraphics[width=11.8cm,clip]{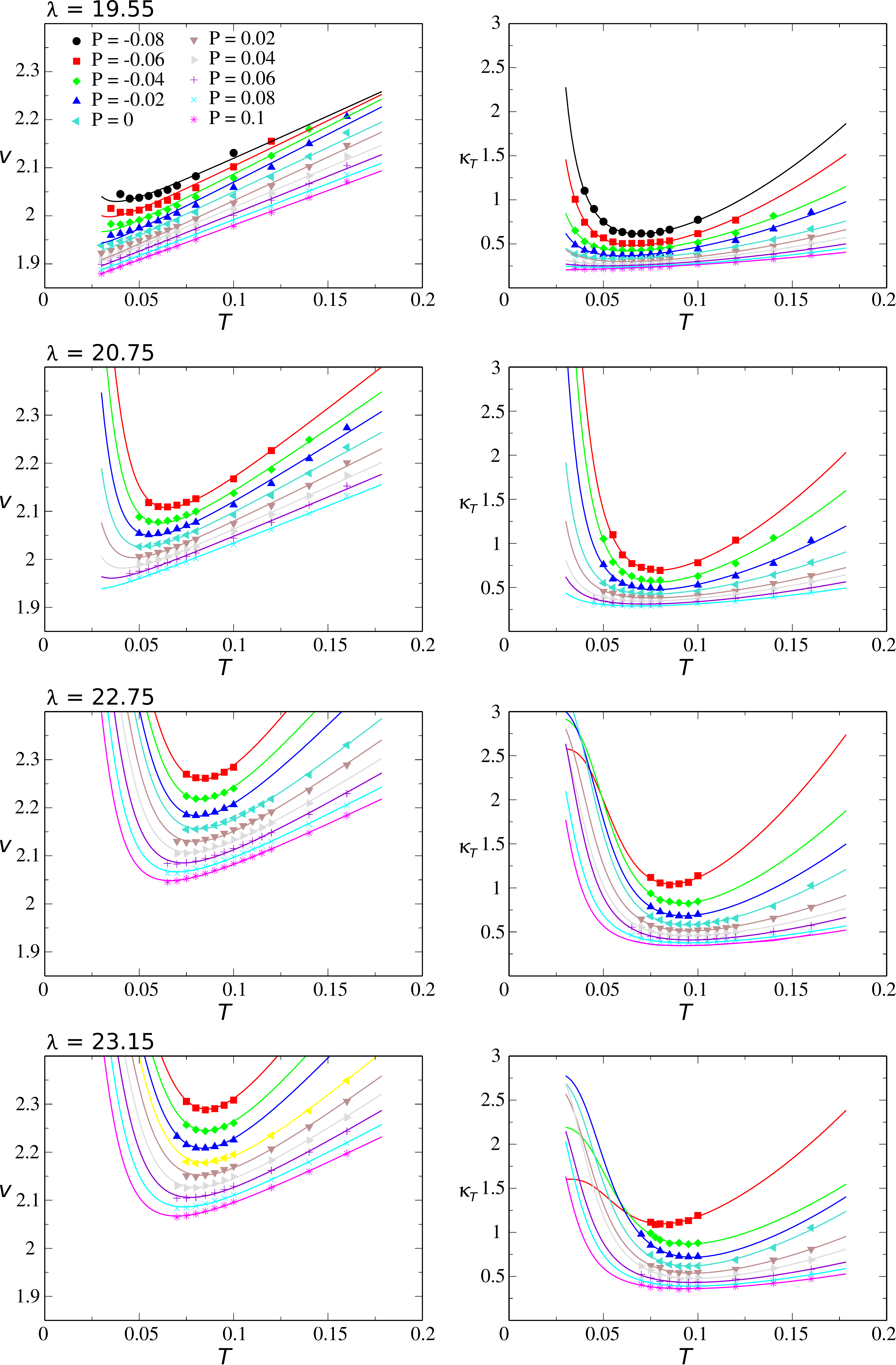}
 \caption{{\bf Thermodynamic anomalies as a function of $\lambda$.} From the top to the bottom row, $\lambda=19.55,20.75,22.75,23.15$. (Left column) Specific volume ($v=1/\rho$) as a function of $T$ and for different $P$. (Right column) Same as left column, but for the isothermal compressibility $\kappa_T$. Symbols are results from simulations, while lines are fits according to the two-state model, Eq.~(\ref{eq:v})-(\ref{eq:k}).}
 \label{fig:thermo_anomalies}
\end{figure*}

We have run extensive computer simulations to map the specific volume and compressibility anomalies in the $(T,P)$ plane, for the values of $\lambda=19.55$, 20.75, 22.75, and 23.15. For each value of $\lambda$ we perform a multiparameter fit, where all simulation results are fitted against Eqs.~(\ref{eq:s}), (\ref{eq:v}), and (\ref{eq:k}), which allows us to obtain the two-state model parameters $\Delta E$, $\Delta v$, and $g$.
In Fig.~\ref{fig:thermo_anomalies} we plot both the density maxima (left column) and compressibility minima (right column) anomalies for $\lambda=19.55$, 20.75, 22.75, and 23.15 (from top to bottom row). All anomalies shift to higher temperature with increasing $\lambda$, while also becoming more pronounced. The two-state model (continuous lines) provides an excellent description of the anomalous behaviour.

\begin{figure*}[!t]
 \centering
 \includegraphics[width=12cm]{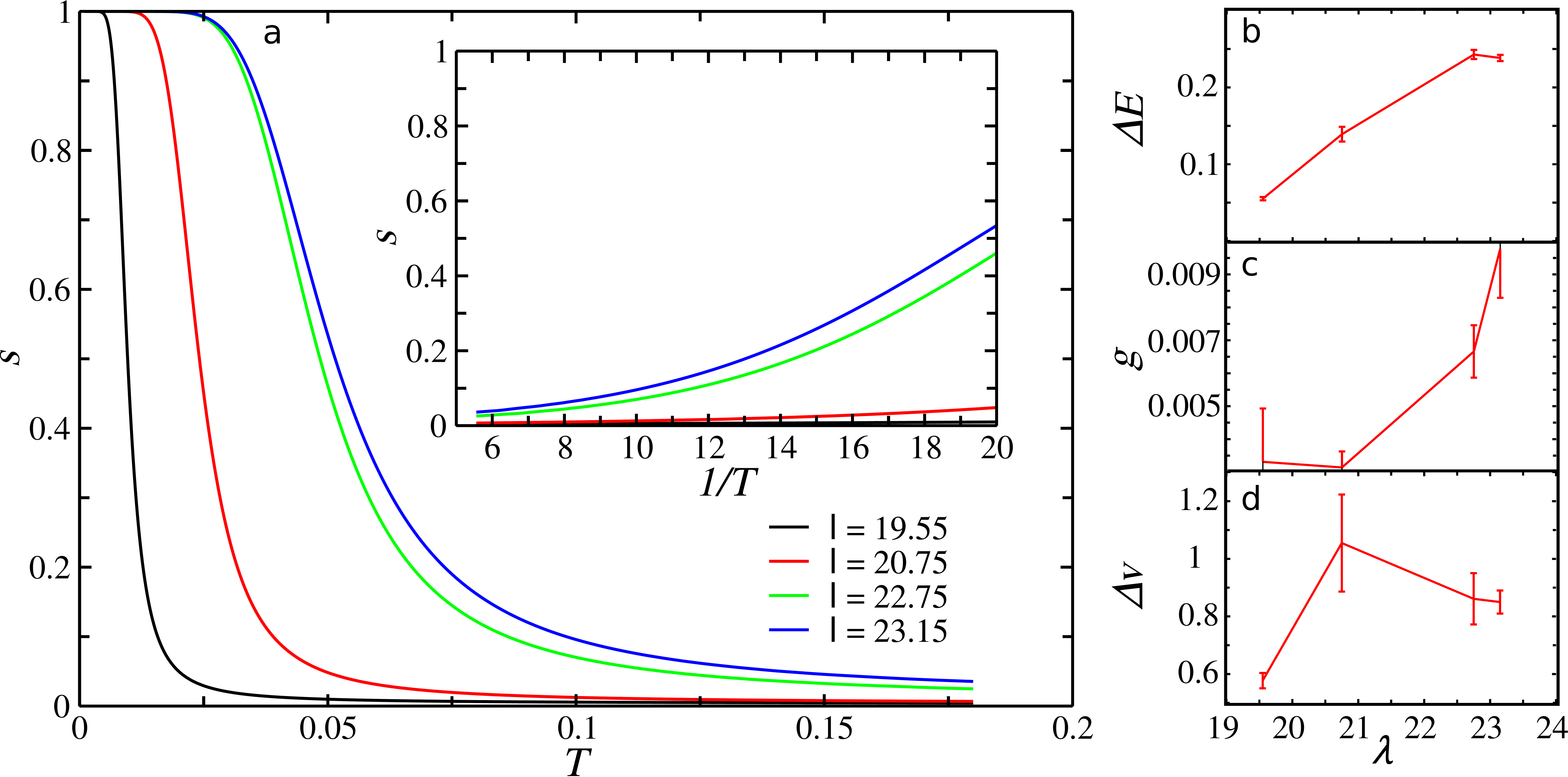}
 \caption{{\bf Two state model anaysis.} {\bf a,} Fraction of $S$-state as a function of $T$ for $P=0$ and different values of $\lambda$. The inset display $s$ as a function of inverse temperature. {\bf b}, {\bf c}, and {\bf d} depict the $\lambda$ dependence of two-state model parameters, $\Delta E$, $g$, and $\Delta v$ respectively.}
 \label{fig:two_state}
\end{figure*}

In Fig.~\ref{fig:two_state} we plot the two-state model parameters obtained by fitting the thermodynamic anomalies of Fig.~\ref{fig:thermo_anomalies}. Figure~\ref{fig:two_state}a shows the increase of the fraction of $S$-state with decreasing $T$, and for different values of $\lambda$. As $\lambda$ is decreased from $\lambda=23.15$ (the value of mW-water), the fraction $s$ decreases, and $T_\text{W}$ (the Widom temperature, where $s=1/2$) moves to lower temperatures. Also the variation of $s$ with $T$ becomes steeper at lower values of $\lambda$, meaning that the anomalies become more localized at lower $T$. To understand these results, in Figs.~\ref{fig:two_state}b-d we plot the variation with $\lambda$ of the parameters $\Delta E$, $g$, and $\Delta v$ respectively. $\Delta E$, the energy difference between the $S$ and $\rho$ states, has the strongest dependence with $\lambda$, increasing by almost a factor of $5$ going from $\lambda=19.55$ to $\lambda=23.15$. Similarly to $\Delta E$, also $g$, the ratio 
between the degeneracies of the $S$ state and $\rho$ state, increases rapidly with $\lambda$. This rapid increase in $g$ is more likely due to a decrease in the degeneracy of the $\rho$ state: as $\lambda$ is increased, the liquid becomes progressively more ordered. Taken together, the increase of both $\Delta E$ and $g$ at high $\lambda$ causes the emergence of anomalous behaviour at higher temperatures, and can be understood as an increase in the tetrahedral ordering of the fluid with $\lambda$ (which controls the strength of the three-body interaction).
They are also responsible for the ease of crystallization of the systems at high $\lambda$, and the high glass-forming ability at low $\lambda$. At lower $\lambda$ the thermodynamic driving force to form locally favoured structures decreases, as the energy gain strongly decreases and the entropy loss also increases. The behaviour of $\Delta v$ in Fig.~\ref{fig:two_state}d is less conclusive, but its decrease at high values of $\lambda$ is in agreement with the change of the slope of the melting line at high $\lambda$ displayed in Fig.~\ref{fig:melting}. The increase in structural order in the $\rho$ state with increasing $\lambda$, which is seen in the $\lambda$-dependence of $g$, may be responsible for the decrease in $\Delta v$.

\subsection*{Dynamic anomalies}

The assumption about the two-state nature of water poses strong constraints on the nature of dynamic anomalies. As explained in the above section of {\it two state model}, our two-state model predicts a \emph{strong-to-strong} transition, contrary to the \emph{fragile-to-strong} transition predicted by scenarios based on the glass transition phenomenology. Here we emphasize that the strong-to-strong transition is 
the Arrhenius-to-Arrhenius transition, and is independent from the glass transition. This is evident from the fact that 
the transition takes place far above the glass transition point ($\sim 2T_g$). 
Thus, the term ``strong'' liquid simply means a simple liquid obeying an Arrhenius law in this context.
On a practical level, the transition from a strong $\rho$-state rich liquid to a strong $S$-state rich liquid can only be followed up to $s\lesssim 0.5$, as crystallization intervenes at high values of $\lambda$, while at low values of $\lambda$ the increase of $s$ is very weak in the observable $T$-window, due to the small energy and large entropy difference (Figs.~\ref{fig:two_state}b and c). 
Note that smaller $\lambda$ means weaker directional bonds, resulting in the smaller energy difference between $\rho$ and $S$ states as well as the weaker constraint on particle configuration for the $\rho$ state,
which leads to the large degeneracy of $\rho$ state.

In order to study the dynamic behaviour as a function of tetrahedrality, we run molecular dynamic simulations covering almost all the accessible region of the $T-\lambda$ parameter space, and keeping $P=0$. The simulations are limited at high $T$ by the location of the liquid-gas spinodal (beyond which there is cavitation), and at low $T$ either by crystallization (for $\lambda\lesssim 18$ and $\lambda\gtrsim 19$) or dynamical slowing down (for $18\lesssim\lambda\lesssim 19$). Simulations are equilibrated in two steps, with isobaric-isothermal Monte Carlo first, and isothermal molecular dynamics second. After equilibration, simulations are conducted in the microcanonical ensemble.

\begin{figure*}[!t]
 \centering
 \includegraphics[width=12cm]{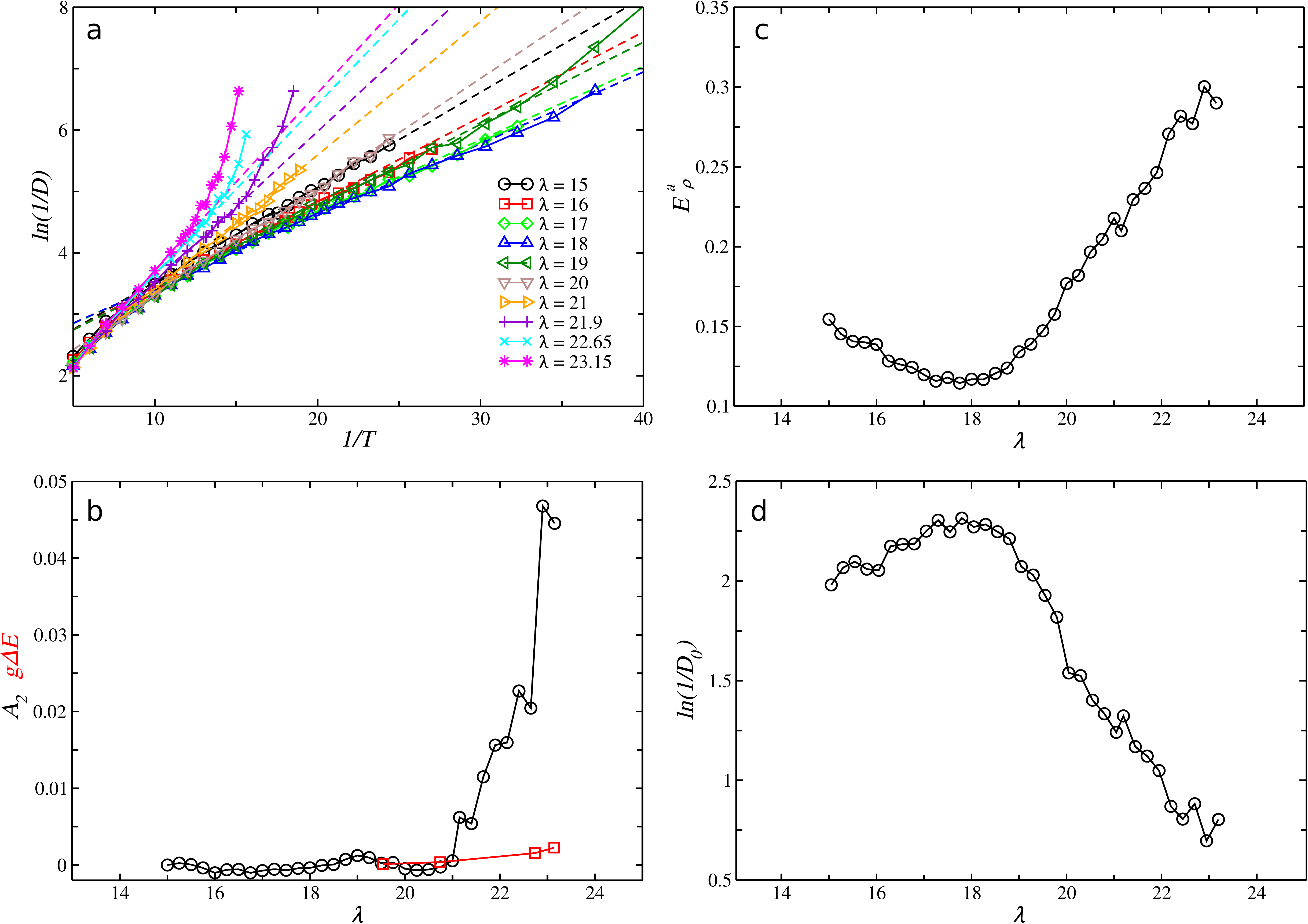}
 \caption{{\bf Dynamics and anomalous behaviour.} {\bf a,} Arrhenius plot for the inverse of the diffusion coefficient $1/D$ vs the inverse temperature $1/T$. Symbols are simulation results, while the dashed lines are the Arrhenius fit to the high-$T$ behaviour. {\bf b,} Coefficient of the quadratic term in Eq.~(\ref{eq:expansion}) (black circle symbols), and the product of $g\Delta E$ (red square symbols). {\bf c,} The activation energy of $\rho$ state, $E^a_\rho$, as a function of $\lambda$. {\bf d,} Inverse of the diffusion prefactor, $1/D_0$, as a function of $\lambda$.}
 \label{fig:dynamic}
\end{figure*}

Figure~\ref{fig:dynamic}a shows the $T$ dependence of the diffusion coefficient for selected values of $\lambda$. Our range goes from very high $T$ ($T_\text{max}=0.2$, which in mW units~\cite{romano2014novel} corresponds to approximately $T_\text{max}=622$~K), down to the homogeneous nucleation temperature. We find that if we include large temperatures, the diffusion coefficient displays sub-Arrhenius behaviour, but if we limit the fits to low temperatures we recover Arrhenius behaviour. Interestingly this is the same behaviour observed in lattice models of two-dimensional doped antiferromagnets without quenched disorder~\cite{kennett2005heterogeneous}. Focusing on the low-temperature behaviour, we note that deviations from the Arrhenius behaviour appear only at high values of $\lambda$, while for $\lambda\lesssim 21$ the relaxation appears to be Arrhenius down to the lowest temperatures. For $\lambda\gtrsim 21$ the behaviour changes from Arrhenius to super-Arrhenius with lowering $T$. We already note that 
this is in contradiction with the glass-transition scenario (see the section of {\it two state model}), which predicts the opposite transition, from super-Arrhenius to Arrhenius (i.e. the \emph{fragile-to-strong} transition).
The observed behaviour is instead fully compatible with the two-state scenario for dynamic anomalies. The inset of Fig.~\ref{fig:two_state}a shows the amount of $S$-state in the same range of $1/T$ where deviations from Arrhenius behaviour appear. For $\lambda\lesssim 21$ the fraction $s$ is negligible, and thus we expect the dynamics to display the strong (Arrhenius) behaviour of the $\rho$ state. For $\lambda\gtrsim 21$, instead, the fraction of $s$ increases considerably, and we thus expect the system to display a transition from the strong (Arrhenius) behaviour of the $\rho$-state to the strong (Arrhenius) behaviour of the $S$-state. According to Eq.~(\ref{eq:expansion}) this transition can be fitted quadratically in $\beta=1/k_{\rm B}T$, and in Fig.~\ref{fig:dynamic}b we plot the quadratic coefficient $A_2$ (black circle symbols) as a function of $\lambda$. The value of $A_2$ confirms that the quadratic term is negligible for $\lambda\lesssim 21$ and increases considerably at higher $\lambda$. The connection of fragile behaviour at $\lambda\gtrsim 21$ with the increase of $S$-state, and the observation of super-Arrhenius behaviour at high $\lambda$ emerging continuously from a pure Arrhenius relaxation at low $\lambda$, strongly supports the two-state interpretation of the dynamic anomaly.

From Eq.~(\ref{eq:expansion}) we know that the quadratic term of the high-$T$ expansion is $A_2=\Delta E^a g(\Delta E-P\Delta v)$, where we can distinguish a dynamical term $\Delta E^a$, which is the difference in the activation energy between the $S$ and $\rho$ state, and a static term $g\Delta E$, where, without loss of generality, we used the fact that we are working at $P=0$. In Fig.~\ref{fig:dynamic}b we superimpose the static term $g\Delta E$ (red square symbols), showing that it has a much weaker $\lambda$ dependence than the quadratic term $A_2$. This implies that also the dynamic term $\Delta E^a$ is a strongly increasing function of $\lambda$ (note that $g$ is a constant). So the effect of tetrahedrality is to increase not only   the energy ($\Delta E$) and entropy difference ($g$) between the $S$ and $\rho$ state, but also the difference in their activation energies, $\Delta E^a$. 
The comparison of $A_2$ and $g\Delta E$ in Fig.~\ref{fig:dynamic}b clearly shows that the increase in $\Delta E^a$ (i.e., the latter) is the main cause of the non-Arrhenius behaviour.
The $\lambda$-dependences of $\Delta E$ and $\Delta E_a$ explain why 
static and dynamic anomalies emerge from ordinary fluid behaviour at high $\lambda$ respectively.

Figures~\ref{fig:dynamic}c and d show how the Arrhenius behaviour of the $\rho$ state changes with $\lambda$. We observe in particular that the activation energy $E_{\rho}^a$ has a minimum around $18\lesssim\lambda\lesssim 19$, which explains why the diffusion constant has a maximum in this region. This minimum in $E_{\rho}^a$ may be a consequence of the competition between density and bond orderings \cite{TanakaWPRB}: Small $\lambda$ ($\lambda \leq 18$) leads to a higher density and weaker directional bonds, whereas 
large $\lambda$ ($\lambda \geq 18$) leads to a lower density and stronger bonds. Note that both higher density and stronger bonds 
result in the higher activation energy. 

\subsection*{Anomalies and apparent divergences}

We have seen that changing tetrahedrality is an effective tool to understand how both thermodynamic and dynamic anomalies emerge from ordinary liquid behaviour. Here we show that altering $\lambda$ can change the behaviour of a water-like liquid  at extreme conditions, and affect its stability limit. We focus in particular on the liquid-gas spinodal line, or more precisely the line of liquid stability, below which the liquid becomes unstable to gas cavitation.
We point out that simulation studies cannot access a true line of instability, as the cavitation of vapour is strongly system-size dependent. We nevertheless use the word ``spinodal'' to refer to this instability, as it is commonly used in the water literature~\cite{dhabal,gonzalez2016comprehensive,angell2016potential}.
In order to determine this line we employ two different procedures.

First, we calculate the density dependence of the inverse of the isothermal compressibility at each temperature, and obtain the spinodal points as the density where the inverse of isothermal compressibility sharply changes. The isothermal compressibility is computed in the NVT ensemble via block analysis~\cite{rovere1988block}, where the distribution of the density order parameter is computed in blocks of different sizes. In the second procedure, which we employ at lower $T$, we run extensive $NVT$ simulations at size $N=512$ at different densities, and equilibrate the equation of state in the unstable region~\cite{block2010curvature,prestipino2015shapes}, where the spinodal point can be obtained from the condition ${\frac{\partial P}{\partial\rho}}|_T=0$ and $\frac{d^2\,P}{d\,\rho^2}>0$. The results of the two methods match in the region of intermediate temperatures.
To get lines of density maxima, we compute the isobaric temperature dependence of densities and obtain temperatures of density maxima by polynomial fitting. The same procedure is applied to compute the line of compressibility minima.

\begin{figure*}[!ht]
 \centering
 \includegraphics[width=12cm]{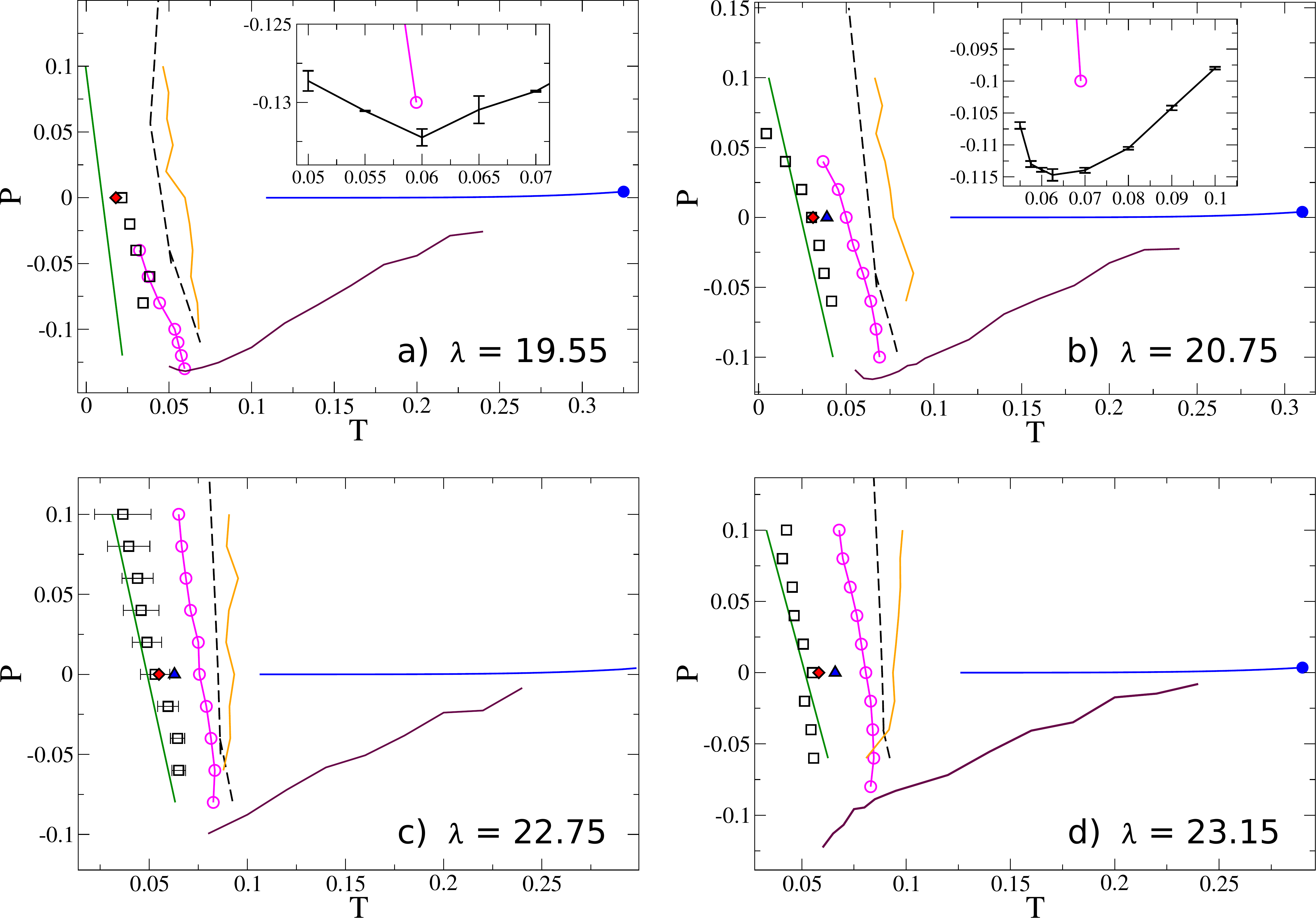}
 \caption{{\bf Stability limits and anomalies.} The different panels differ for their value of $\lambda$: $19.55$ ({\bf a}), $20.75$ ({\bf b}), $22.75$ ({\bf c}), and $23.15$ ({\bf d}). In each panel: blue full circle is the liquid-gas critical point, the continuous blue line is the liquid-gas coexistence line, the continuous purple line is the liquid spinodal (or better stability limit of the liquid phase), the orange continuous line is the line of compressibility minima, the open magenta symbols are the line line of density maxima, the open square symbols are the location of the spinodal line as extrapolated from the apparent divergence of the compressibility, which may or may not be contiguous with the liquid spinodal, the red diamond full symbol is the ideal glass transition temperature as extrapolated by VFT fit of the diffusivity, the blue triangle full symbol is the homogeneous nucleation point at $P=0$, the continuous green line is the Schottky (or Widom) line as predicted by the two-state model, and the dashed black lines represent the liquid-solid coexistence line.}
 \label{fig:summary}
\end{figure*}

In Fig.~\ref{fig:summary} we summarize the loci of thermodynamic anomalies and liquid stability for $\lambda=19.55$ (a), $20.75$ (b), $22.75$ (c), and $23.15$ (d). The most notable change occurs to the liquid spinodal (continuous purple line) that emerges from the liquid-gas critical point (full blue circle symbol): while at high values of $\lambda$ (panels c and d) the spinodal displays usual monotonous behaviour, for low $\lambda$ (panels a and b) the spinodal intersects the line of density maxima (purple open circle symbols) and retraces.
Comparing these results with Ref.~\cite{sastry_ll}, where the line of density maxima for Silicon ($\lambda=21$) was shown to just miss the spinodal line, we can estimate that the re-entrant behaviour of the spinodal starts at approximately $\lambda<21$.
To our knowledge this is the first microscopic model of water-like liquids that displays a transition to a re-entrant spinodal~\cite{speedy1982stability}, a result which was predicted in terms of a mean-field cell model~\cite{stokely2010effect}. Very recently, in Patchy Particles' colloidal models, the authors of Ref.~\cite{lorenzo_spinodal} also observed a retracing spinodal, which in their case extended to positive pressures.

Next we focus on the apparent divergences in thermodynamic and dynamic properties of water at low $T$. In Fig.~\ref{fig:summary} we plot as open square symbols the estimated location of the apparent spinodal divergence $T_{\rm sp}(P)$, as obtained from fitting the increase of the isothermal compressibility with the following relation
\begin{equation}
 \kappa_T(T,P)=k(P)T^2+n(P)+K(P)(T-T_{\rm sp} (P))^{-\gamma},
\end{equation}
where the first two terms are the background behaviour of the compressibility, and whose coefficients are the same as the ones employed in Eq.~(\ref{eq:k}). We also plot as a red diamond full symbol the apparent dynamic divergence at $P=0$, as obtained by a Vogel–Fulcher–Tammann (VFT) fit of the diffusion data (Fig.~\ref{fig:dynamic}a). Finally, we also plot the location of the Widom line (green continuous line) as obtained from the two-state model (Eq.~(\ref{eq:widom})), which is not a divergence, but the line along which $s=1/2$. 
The spinodal line, the glass transition point, and the Widom line cannot be accessed in equilibrium, as they lie below the homogeneous nucleation line, whose $P=0$ point is plotted as a blue full triangle symbol. It is important to observe that the spinodal divergence and the dynamical divergence fall on top of each other within the errors, and on top of the Widom line. We have shown that the spinodal divergence should not occur at high $\lambda$ (panels c and d), where there is no retracing spinodal, and also the dynamic divergence does not occur at low values of $\lambda$, where the relaxation is more consistently fitted as Arrhenius (see Fig.~\ref{fig:dynamic}). These observations strongly hint to the fact that these divergences are only apparent. Their coincidence with the Widom (or Schottky) line, indicates that the apparent divergences simply point to the loci of maximum change in the behaviour of water, as predicted as the Schottky anomaly  of the two-state model: from the $\rho$ state to the $S$ state in the case of thermodynamic anomalies, and from the high $T$ Arrhenius (strong) regime, to the low $T$ Arrhenius (strong) regime. We thus believe that the two-state model can rationalize all the observation of both thermodynamic and dynamic behaviour across all values of $\lambda$, interpolating between simple liquid behaviour found at low $\lambda$, and the rich interplay of anomalies found at high $\lambda$.

\section*{Discussion}
In this Article we have exploited the strategy of varying the tetrahedrality of the SW model in order to gain insights into the anomalous behaviour of water and other tetrahedral materials in their liquid state as well as the phase behaviour including 
all gas, liquid, and solid phases.
The SW model has found widespread applicability in the study of thermodynamic anomalies of tetrahedral liquids,
most notably silicon~\cite{sastry_ll} and water~\cite{limmer_molinero}. The first study to consider variations of $\lambda$ as a means to change continuously the property of the materials was the seminal study of Ref.~\cite{molinero2006tuning}, where the glass forming ability was considered. Very recently the same idea was also applied to study the change in the anomalous properties of the liquid phase~\cite{angell2016potential,dhabal}. 

In our work, we have computed the full phase diagram of the SW model as a function of the tetrahedral parameter $\lambda$. For the first time, we have determined the phase diagram at negative pressures, and also computed the $\lambda$ dependence of the critical point. We then focused on liquid anomalies, with a special focus on the negative pressure region.



To rationalize the behaviour of the anomalies we have then applied a two-state model, fitting both the density and compressibility anomalies. The two-state model predicts an increase of the driving force towards the more ordered $S$-state with increasing $\lambda$: both the difference in energy $\Delta E$ and in degeneracy $g$ increase considerably with $\lambda$, as a consequence of a tendency from the liquid to become more ordered locally as the strength of the tetrahedral interaction becomes stronger. This explains how the anomalies emerge continuously by increasing $\lambda$, moving progressively towards lower $T$ and $P$.

We have then analyzed the behaviour of dynamic anomalies, focusing on diffusion. We  have shown that at small $\lambda$ the dynamics is Arrhenius, while at large $\lambda$ the dynamics crosses to super-Arrhenius. The emergence of super-Arrhenius behaviour from Arrhenius behaviour, in coincidence with the increase in the fraction of $S$-states, is in line with the predictions of the two-state model, i.e., strong (Arrhenius)-to-strong (Arrhenius) transition, while it is at odds with interpretations based on the glass transition singularity, i.e., fragile-to-strong transition. From a quadratic fit of the $T$ dependence of the diffusion coefficient, we  have also found that the activation energy difference $\Delta E^a_\rho$ is a strongly increasing function of $\lambda$.

We have also considered the location of the anomalies and apparent divergences in relation to the phase behaviour. We have found that by lowering $\lambda$ the phase diagram changes to a retracing spinodal scenario, which occurs when the line of density maxima crosses the spinodal line.
Increasing $\lambda$, the landscape changes from a retracing spinodal to a monotonous one, and the dynamic relaxation changes from Arrhenius to apparently super-Arrhenius. Despite these changes, all extrapolations based on singular behaviour (spinodal divergences for thermodynamic anomalies, and glass divergence for dynamic ones), always fall on top of the predicted two-state Schottky (or Widom) line. 
Starting from locally favoured structures (the $S$-state), the two-state model provides a unified description of water anomalies that is independent of singularities, while still being compatible with them.

Finally, our study reveals that water is the material where tetrahedrality plays the bigger role: 
if tetrahedrality is weaker than that of water, the two-state feature becomes weaker, while if it is stronger than water, on the other hand, 
the volume difference between the two states becomes smaller, leading to a weaker density anomaly. 
On noting that the two-state feature is the origin of the flexibility of water properties, or the large susceptibility of the properties to 
physical and chemical perturbations, our finding highlights the exceptional nature of water, which makes it so special compared to any other substances.

\vspace{2mm}
\noindent
{\bf Acknowledgments.}
This work was partially supported by Grants-in-Aid for Specially Promoted Research (25000002) from the Japan Society of the Promotion of Science (JSPS).
JR acknowledges support from the ERC Grant DLV-759187 and the Royal Society University Research Fellowship.

\section*{Materials and Methods}
\small

\subsection{SW potential}
Here, the pairwise term $U_2$ models a steep repulsion at short distances and a short-range attraction,
\begin{equation}\label{eq:sw2}
U_2(r)=A\epsilon\left[B\left(\frac{\sigma}{r}\right)^{p}-\left(\frac{\sigma}{r}\right)^{q}\right]\mathrm{exp}\left(\frac{\sigma}{r-a\sigma}\right), \nonumber \\
\end{equation}
while the three-body interaction term $U_3$ is a directional repulsive interaction which promotes tetrahedral angles between triplets of particles,
\begin{equation}
\begin{split}
U_3(r_{ij},r_{ik})=&\epsilon[\cos\theta_{ijk}-\cos\theta_{0}]^2 \times\\
&\exp \left(\frac{\gamma\sigma}{r_{ij}-a\sigma}\right)\exp \left(\frac{\gamma\sigma}{r_{ik}-a\sigma}\right). \nonumber
\end{split}
\end{equation}
The parameters for the models in this work are $A=7.049556277$, $B=0.6022245584$, $p=4$, $q=0$, $\cos\theta_{0}=-1/3$, $\gamma=1.2$, and $a=1.8$.
The parameter $\epsilon$ sets the energy scale and $\sigma$ the length scale.
They correspond to the depth of the two-body interaction potential and the particle diameter respectively, and determined by materials for which the model is used.
We use internal units where $\epsilon$ and $\sigma$ are the units of energy and length respectively.

\vspace{5mm}

\subsection{Numerical methods}
In order to compute solid-liquid and liquid-gas coexistence lines, we run Monte Carlo simulations in the isothermal-isobaric $NPT$ ensembles.
The size and shape of the simulation box can fluctuate so as to allow crystalline phases to change their structures \cite{boxshape1,boxshape2}.
A volume-change attempt occurs every $N$ translation attempts.
The number of particles in the box is $N=1024$.
We perform Gibbs-Duhem integration~\cite{kofke} and Hamiltonian Gibbs-Duhem integration~\cite{Vega} in order to obtain coexistence lines along the pressure axis and along $\lambda$ axis respectively.
Triple lines are computed in the same way as in Ref.~\cite{akahane_quadruple}.

In order to obtain liquid-gas critical points, we run Monte Carlo simulations in the grand canonical ensemble. We compute the distribution functions of the mixing order parameter $M$ ($M=\rho+m u$; $\rho$ is density and $u$ is internal energy per particle, and 
$m$ is mixing parameter), and use histogram re-weighting methods~\cite{reweight} to fit them into the Ising universal curve~\cite{tsypin}. Liquid-gas coexistence lines are instead computed by locating a coexistence point close to the critical point with Successive Umbrella Sampling simulations, and then running Gibbs-Duhem integration to trace the coexistence line at lower temperatures.

To compute liquid-gas spinodal points, we follow two strategies. In the first strategy we compute isothermal compressibilities dividing the simulation box in smaller boxes to evaluate the size dependence
of the compressibility; we then define the spinodal points as where the inverse of the compressibility vanishes. At lower temperatures, we instead run simulations in the $NVT$ ensemble and constructed the whole equation of state~\cite{block2010curvature,prestipino2015shapes}, detecting the spinodal points as the points where ${\frac{\partial P}{\partial\rho}}|_T=0$. In order to equilibrate simulations in the unstable region, we reduced the number of particles to $N=512$. Both techniques gave similar results in the region of overlap.
To obtain lines of density maxima and compressibility minima, we run $NPT$ Monte Carlo simulations and compute averages and fluctuations of densities.

\normalsize



\begin{thebibliography}{75}%
\makeatletter
\providecommand \@ifxundefined [1]{%
 \@ifx{#1\undefined}
}%
\providecommand \@ifnum [1]{%
 \ifnum #1\expandafter \@firstoftwo
 \else \expandafter \@secondoftwo
 \fi
}%
\providecommand \@ifx [1]{%
 \ifx #1\expandafter \@firstoftwo
 \else \expandafter \@secondoftwo
 \fi
}%
\providecommand \natexlab [1]{#1}%
\providecommand \enquote  [1]{``#1''}%
\providecommand \bibnamefont  [1]{#1}%
\providecommand \bibfnamefont [1]{#1}%
\providecommand \citenamefont [1]{#1}%
\providecommand \href@noop [0]{\@secondoftwo}%
\providecommand \href [0]{\begingroup \@sanitize@url \@href}%
\providecommand \@href[1]{\@@startlink{#1}\@@href}%
\providecommand \@@href[1]{\endgroup#1\@@endlink}%
\providecommand \@sanitize@url [0]{\catcode `\\12\catcode `\$12\catcode
  `\&12\catcode `\#12\catcode `\^12\catcode `\_12\catcode `\%12\relax}%
\providecommand \@@startlink[1]{}%
\providecommand \@@endlink[0]{}%
\providecommand \url  [0]{\begingroup\@sanitize@url \@url }%
\providecommand \@url [1]{\endgroup\@href {#1}{\urlprefix }}%
\providecommand \urlprefix  [0]{URL }%
\providecommand \Eprint [0]{\href }%
\providecommand \doibase [0]{http://dx.doi.org/}%
\providecommand \selectlanguage [0]{\@gobble}%
\providecommand \bibinfo  [0]{\@secondoftwo}%
\providecommand \bibfield  [0]{\@secondoftwo}%
\providecommand \translation [1]{[#1]}%
\providecommand \BibitemOpen [0]{}%
\providecommand \bibitemStop [0]{}%
\providecommand \bibitemNoStop [0]{.\EOS\space}%
\providecommand \EOS [0]{\spacefactor3000\relax}%
\providecommand \BibitemShut  [1]{\csname bibitem#1\endcsname}%
\let\auto@bib@innerbib\@empty
\bibitem [{\citenamefont {Debenedetti}(1996)}]{debenedetti1996metastable}%
  \BibitemOpen
  \bibfield  {author} {\bibinfo {author} {\bibfnamefont {Pablo~G}\ \bibnamefont
  {Debenedetti}},\ }\href@noop {} {\emph {\bibinfo {title} {Metastable liquids:
  concepts and principles}}}\ (\bibinfo  {publisher} {Princeton University
  Press},\ \bibinfo {year} {1996})\BibitemShut {NoStop}%
\bibitem [{\citenamefont {Mishima}\ and\ \citenamefont
  {Stanley}(1998)}]{Mishima1998}%
  \BibitemOpen
  \bibfield  {author} {\bibinfo {author} {\bibfnamefont {Osamu}\ \bibnamefont
  {Mishima}}\ and\ \bibinfo {author} {\bibfnamefont {H.~Eugene}\ \bibnamefont
  {Stanley}},\ }\bibfield  {title} {\enquote {\bibinfo {title} {{The
  relationship between liquid, supercooled and glassy water}},}\ }\href
  {\doibase 10.1038/24540} {\bibfield  {journal} {\bibinfo  {journal} {Nature}\
  }\textbf {\bibinfo {volume} {396}},\ \bibinfo {pages} {329--335} (\bibinfo
  {year} {1998})}\BibitemShut {NoStop}%
\bibitem [{\citenamefont {Debenedetti}(2003)}]{debenedetti2003supercooled}%
  \BibitemOpen
  \bibfield  {author} {\bibinfo {author} {\bibfnamefont {Pablo~G}\ \bibnamefont
  {Debenedetti}},\ }\bibfield  {title} {\enquote {\bibinfo {title} {Supercooled
  and glassy water},}\ }\href@noop {} {\bibfield  {journal} {\bibinfo
  {journal} {J. Phys.: Condens. Matter}\ }\textbf {\bibinfo {volume} {15}},\
  \bibinfo {pages} {R1669} (\bibinfo {year} {2003})}\BibitemShut {NoStop}%
\bibitem [{\citenamefont {Nilsson}\ and\ \citenamefont
  {Pettersson}(2015)}]{Nilsson2015}%
  \BibitemOpen
  \bibfield  {author} {\bibinfo {author} {\bibfnamefont {Anders}\ \bibnamefont
  {Nilsson}}\ and\ \bibinfo {author} {\bibfnamefont {Lars G.~M.}\ \bibnamefont
  {Pettersson}},\ }\bibfield  {title} {\enquote {\bibinfo {title} {{The
  structural origin of anomalous properties of liquid water.}}}\ }\href@noop {}
  {\bibfield  {journal} {\bibinfo  {journal} {Nat. Commun.}\ }\textbf {\bibinfo
  {volume} {6}},\ \bibinfo {pages} {8998} (\bibinfo {year} {2015})}\BibitemShut
  {NoStop}%
\bibitem [{\citenamefont {Gallo}\ \emph {et~al.}(2016)\citenamefont {Gallo},
  \citenamefont {Amann-Winkel}, \citenamefont {Angell}, \citenamefont
  {Anisimov}, \citenamefont {Caupin}, \citenamefont {Chakravarty},
  \citenamefont {Lascaris}, \citenamefont {Loerting}, \citenamefont
  {Panagiotopoulos}, \citenamefont {Russo}, \citenamefont {Sellberg},
  \citenamefont {Stanley}, \citenamefont {Tanaka}, \citenamefont {Vega},
  \citenamefont {Xu},\ and\ \citenamefont {Pettersson}}]{gallo2016water}%
  \BibitemOpen
  \bibfield  {author} {\bibinfo {author} {\bibfnamefont {Paola}\ \bibnamefont
  {Gallo}}, \bibinfo {author} {\bibfnamefont {Katrin}\ \bibnamefont
  {Amann-Winkel}}, \bibinfo {author} {\bibfnamefont {Charles~Austen}\
  \bibnamefont {Angell}}, \bibinfo {author} {\bibfnamefont
  {Mikhail~Alexeevich}\ \bibnamefont {Anisimov}}, \bibinfo {author}
  {\bibfnamefont {Fre?de?ric}\ \bibnamefont {Caupin}}, \bibinfo {author}
  {\bibfnamefont {Charusita}\ \bibnamefont {Chakravarty}}, \bibinfo {author}
  {\bibfnamefont {Erik}\ \bibnamefont {Lascaris}}, \bibinfo {author}
  {\bibfnamefont {Thomas}\ \bibnamefont {Loerting}}, \bibinfo {author}
  {\bibfnamefont {Athanassios~Zois}\ \bibnamefont {Panagiotopoulos}}, \bibinfo
  {author} {\bibfnamefont {John}\ \bibnamefont {Russo}}, \bibinfo {author}
  {\bibfnamefont {Jonas~Alexander}\ \bibnamefont {Sellberg}}, \bibinfo {author}
  {\bibfnamefont {Harry~Eugene}\ \bibnamefont {Stanley}}, \bibinfo {author}
  {\bibfnamefont {Hajime}\ \bibnamefont {Tanaka}}, \bibinfo {author}
  {\bibfnamefont {Carlos}\ \bibnamefont {Vega}}, \bibinfo {author}
  {\bibfnamefont {Limei}\ \bibnamefont {Xu}}, \ and\ \bibinfo {author}
  {\bibfnamefont {Lars Gunnar~Moody}\ \bibnamefont {Pettersson}},\ }\bibfield
  {title} {\enquote {\bibinfo {title} {Water: A tale of two liquids},}\
  }\href@noop {} {\bibfield  {journal} {\bibinfo  {journal} {Chem. Rev.}\
  }\textbf {\bibinfo {volume} {116}},\ \bibinfo {pages} {7463--7500} (\bibinfo
  {year} {2016})}\BibitemShut {NoStop}%
\bibitem [{\citenamefont {Tanaka}(2002)}]{TanakaWPRB}%
  \BibitemOpen
  \bibfield  {author} {\bibinfo {author} {\bibfnamefont {H.}~\bibnamefont
  {Tanaka}},\ }\bibfield  {title} {\enquote {\bibinfo {title} {Simple view of
  waterlike anomalies of atomic liquids with directional bonding},}\
  }\href@noop {} {\bibfield  {journal} {\bibinfo  {journal} {Phys. Rev. B}\
  }\textbf {\bibinfo {volume} {66}},\ \bibinfo {pages} {064202} (\bibinfo
  {year} {2002})}\BibitemShut {NoStop}%
\bibitem [{\citenamefont {Stillinger}\ and\ \citenamefont
  {Weber}(1985)}]{stillingerweber}%
  \BibitemOpen
  \bibfield  {author} {\bibinfo {author} {\bibfnamefont {Frank~H.}\
  \bibnamefont {Stillinger}}\ and\ \bibinfo {author} {\bibfnamefont
  {Thomas~A.}\ \bibnamefont {Weber}},\ }\bibfield  {title} {\enquote {\bibinfo
  {title} {Computer simulation of local order in condensed phases of
  silicon},}\ }\href@noop {} {\bibfield  {journal} {\bibinfo  {journal} {Phys.
  Rev B}\ }\textbf {\bibinfo {volume} {31}},\ \bibinfo {pages} {5262} (\bibinfo
  {year} {1985})}\BibitemShut {NoStop}%
\bibitem [{\citenamefont {Molinero}\ and\ \citenamefont
  {Moore}(2008)}]{molinero2008water}%
  \BibitemOpen
  \bibfield  {author} {\bibinfo {author} {\bibfnamefont {Valeria}\ \bibnamefont
  {Molinero}}\ and\ \bibinfo {author} {\bibfnamefont {Emily~B.}\ \bibnamefont
  {Moore}},\ }\bibfield  {title} {\enquote {\bibinfo {title} {Water modeled as
  an intermediate element between carbon and silicon},}\ }\href@noop {}
  {\bibfield  {journal} {\bibinfo  {journal} {J. Phys. Chem. B}\ }\textbf
  {\bibinfo {volume} {113}},\ \bibinfo {pages} {4008--4016} (\bibinfo {year}
  {2008})}\BibitemShut {NoStop}%
\bibitem [{\citenamefont {Moore}\ and\ \citenamefont
  {Molinero}(2011)}]{molinero_nature}%
  \BibitemOpen
  \bibfield  {author} {\bibinfo {author} {\bibfnamefont {E.B.}\ \bibnamefont
  {Moore}}\ and\ \bibinfo {author} {\bibfnamefont {V.}~\bibnamefont
  {Molinero}},\ }\bibfield  {title} {\enquote {\bibinfo {title} {Structural
  transformation in supercooled water controls the crystallization rate of
  ice},}\ }\href@noop {} {\bibfield  {journal} {\bibinfo  {journal} {Nature}\
  }\textbf {\bibinfo {volume} {479}},\ \bibinfo {pages} {506--508} (\bibinfo
  {year} {2011})}\BibitemShut {NoStop}%
\bibitem [{\citenamefont {Reinhardt}\ and\ \citenamefont
  {Doye}(2012)}]{reinhardt2012free}%
  \BibitemOpen
  \bibfield  {author} {\bibinfo {author} {\bibfnamefont {Aleks}\ \bibnamefont
  {Reinhardt}}\ and\ \bibinfo {author} {\bibfnamefont {Jonathan~PK}\
  \bibnamefont {Doye}},\ }\bibfield  {title} {\enquote {\bibinfo {title} {Free
  energy landscapes for homogeneous nucleation of ice for a monatomic water
  model},}\ }\href@noop {} {\bibfield  {journal} {\bibinfo  {journal} {J. Chem.
  Phys.}\ }\textbf {\bibinfo {volume} {136}},\ \bibinfo {pages} {054501}
  (\bibinfo {year} {2012})}\BibitemShut {NoStop}%
\bibitem [{\citenamefont {Li}\ \emph {et~al.}(2013)\citenamefont {Li},
  \citenamefont {Donadio},\ and\ \citenamefont {Galli}}]{li2013ice}%
  \BibitemOpen
  \bibfield  {author} {\bibinfo {author} {\bibfnamefont {Tianshu}\ \bibnamefont
  {Li}}, \bibinfo {author} {\bibfnamefont {Davide}\ \bibnamefont {Donadio}}, \
  and\ \bibinfo {author} {\bibfnamefont {Giulia}\ \bibnamefont {Galli}},\
  }\bibfield  {title} {\enquote {\bibinfo {title} {Ice nucleation at the
  nanoscale probes no man's land of water},}\ }\href@noop {} {\bibfield
  {journal} {\bibinfo  {journal} {Nature Commun.}\ }\textbf {\bibinfo {volume}
  {4}},\ \bibinfo {pages} {1887} (\bibinfo {year} {2013})}\BibitemShut
  {NoStop}%
\bibitem [{\citenamefont {Russo}\ \emph {et~al.}(2014)\citenamefont {Russo},
  \citenamefont {Romano},\ and\ \citenamefont {Tanaka}}]{russo2014new}%
  \BibitemOpen
  \bibfield  {author} {\bibinfo {author} {\bibfnamefont {John}\ \bibnamefont
  {Russo}}, \bibinfo {author} {\bibfnamefont {Flavio}\ \bibnamefont {Romano}},
  \ and\ \bibinfo {author} {\bibfnamefont {Hajime}\ \bibnamefont {Tanaka}},\
  }\bibfield  {title} {\enquote {\bibinfo {title} {New metastable form of ice
  and its role in the homogeneous crystallization of water},}\ }\href@noop {}
  {\bibfield  {journal} {\bibinfo  {journal} {Nature Mater.}\ }\textbf
  {\bibinfo {volume} {13}},\ \bibinfo {pages} {733--739} (\bibinfo {year}
  {2014})}\BibitemShut {NoStop}%
\bibitem [{\citenamefont {Sosso}\ \emph
  {et~al.}(2016{\natexlab{a}})\citenamefont {Sosso}, \citenamefont {Chen},
  \citenamefont {Cox}, \citenamefont {Fitzner}, \citenamefont {Pedevilla},
  \citenamefont {Zen},\ and\ \citenamefont {Michaelides}}]{sosso2016crystal}%
  \BibitemOpen
  \bibfield  {author} {\bibinfo {author} {\bibfnamefont {Gabriele~C}\
  \bibnamefont {Sosso}}, \bibinfo {author} {\bibfnamefont {Ji}~\bibnamefont
  {Chen}}, \bibinfo {author} {\bibfnamefont {Stephen~J}\ \bibnamefont {Cox}},
  \bibinfo {author} {\bibfnamefont {Martin}\ \bibnamefont {Fitzner}}, \bibinfo
  {author} {\bibfnamefont {Philipp}\ \bibnamefont {Pedevilla}}, \bibinfo
  {author} {\bibfnamefont {Andrea}\ \bibnamefont {Zen}}, \ and\ \bibinfo
  {author} {\bibfnamefont {Angelos}\ \bibnamefont {Michaelides}},\ }\bibfield
  {title} {\enquote {\bibinfo {title} {Crystal nucleation in liquids: Open
  questions and future challenges in molecular dynamics simulations},}\
  }\href@noop {} {\bibfield  {journal} {\bibinfo  {journal} {Chem. Rev.}\
  }\textbf {\bibinfo {volume} {116}},\ \bibinfo {pages} {7078--7116} (\bibinfo
  {year} {2016}{\natexlab{a}})}\BibitemShut {NoStop}%
\bibitem [{\citenamefont {Sosso}\ \emph
  {et~al.}(2016{\natexlab{b}})\citenamefont {Sosso}, \citenamefont {Li},
  \citenamefont {Donadio}, \citenamefont {Tribello},\ and\ \citenamefont
  {Michaelides}}]{sosso2016microscopic}%
  \BibitemOpen
  \bibfield  {author} {\bibinfo {author} {\bibfnamefont {Gabriele~C}\
  \bibnamefont {Sosso}}, \bibinfo {author} {\bibfnamefont {Tianshu}\
  \bibnamefont {Li}}, \bibinfo {author} {\bibfnamefont {Davide}\ \bibnamefont
  {Donadio}}, \bibinfo {author} {\bibfnamefont {Gareth~A}\ \bibnamefont
  {Tribello}}, \ and\ \bibinfo {author} {\bibfnamefont {Angelos}\ \bibnamefont
  {Michaelides}},\ }\bibfield  {title} {\enquote {\bibinfo {title} {Microscopic
  mechanism and kinetics of ice formation at complex interfaces: Zooming in on
  kaolinite},}\ }\href@noop {} {\bibfield  {journal} {\bibinfo  {journal} {J.
  Phys. Chem. Lett.}\ }\textbf {\bibinfo {volume} {7}},\ \bibinfo {pages}
  {2350--2355} (\bibinfo {year} {2016}{\natexlab{b}})}\BibitemShut {NoStop}%
\bibitem [{\citenamefont {Pipolo}\ \emph {et~al.}(2017)\citenamefont {Pipolo},
  \citenamefont {Salanne}, \citenamefont {Ferlat}, \citenamefont {Klotz},
  \citenamefont {Saitta},\ and\ \citenamefont
  {Pietrucci}}]{pipolo2017navigating}%
  \BibitemOpen
  \bibfield  {author} {\bibinfo {author} {\bibfnamefont {Silvio}\ \bibnamefont
  {Pipolo}}, \bibinfo {author} {\bibfnamefont {Mathieu}\ \bibnamefont
  {Salanne}}, \bibinfo {author} {\bibfnamefont {Guillaume}\ \bibnamefont
  {Ferlat}}, \bibinfo {author} {\bibfnamefont {Stefan}\ \bibnamefont {Klotz}},
  \bibinfo {author} {\bibfnamefont {A~Marco}\ \bibnamefont {Saitta}}, \ and\
  \bibinfo {author} {\bibfnamefont {Fabio}\ \bibnamefont {Pietrucci}},\
  }\bibfield  {title} {\enquote {\bibinfo {title} {Navigating at will on the
  water phase diagram},}\ }\href@noop {} {\bibfield  {journal} {\bibinfo
  {journal} {arXiv preprint arXiv:1703.00753}\ } (\bibinfo {year}
  {2017})}\BibitemShut {NoStop}%
\bibitem [{\citenamefont {Holten}\ \emph
  {et~al.}(2013{\natexlab{a}})\citenamefont {Holten}, \citenamefont {Limmer},
  \citenamefont {Molinero},\ and\ \citenamefont {Anisimov}}]{limmer_molinero}%
  \BibitemOpen
  \bibfield  {author} {\bibinfo {author} {\bibfnamefont {Vincent}\ \bibnamefont
  {Holten}}, \bibinfo {author} {\bibfnamefont {David~T.}\ \bibnamefont
  {Limmer}}, \bibinfo {author} {\bibfnamefont {Valeria}\ \bibnamefont
  {Molinero}}, \ and\ \bibinfo {author} {\bibfnamefont {Mikhail~A.}\
  \bibnamefont {Anisimov}},\ }\bibfield  {title} {\enquote {\bibinfo {title}
  {Nature of the anomalies in the supercooled liquid state of the mw model of
  water},}\ }\href {\doibase 10.1063/1.4802992} {\bibfield  {journal} {\bibinfo
   {journal} {J. Chem. Phys.}\ }\textbf {\bibinfo {volume} {138}},\ \bibinfo
  {eid} {174501} (\bibinfo {year} {2013}{\natexlab{a}})}\BibitemShut {NoStop}%
\bibitem [{\citenamefont {Sengupta}\ \emph {et~al.}(2014)\citenamefont
  {Sengupta}, \citenamefont {Vasisht},\ and\ \citenamefont
  {Sastry}}]{sengupta2014diffusivity}%
  \BibitemOpen
  \bibfield  {author} {\bibinfo {author} {\bibfnamefont {Shiladitya}\
  \bibnamefont {Sengupta}}, \bibinfo {author} {\bibfnamefont {Vishwas~V}\
  \bibnamefont {Vasisht}}, \ and\ \bibinfo {author} {\bibfnamefont {Srikanth}\
  \bibnamefont {Sastry}},\ }\bibfield  {title} {\enquote {\bibinfo {title}
  {Diffusivity anomaly in modified stillinger-weber liquids},}\ }\href@noop {}
  {\bibfield  {journal} {\bibinfo  {journal} {J. Chem. Phys.}\ }\textbf
  {\bibinfo {volume} {140}},\ \bibinfo {pages} {044503} (\bibinfo {year}
  {2014})}\BibitemShut {NoStop}%
\bibitem [{\citenamefont {Singh}\ \emph {et~al.}(2014)\citenamefont {Singh},
  \citenamefont {Dhabal}, \citenamefont {Nguyen}, \citenamefont {Molinero},\
  and\ \citenamefont {Chakravarty}}]{singh2014triplet}%
  \BibitemOpen
  \bibfield  {author} {\bibinfo {author} {\bibfnamefont {Murari}\ \bibnamefont
  {Singh}}, \bibinfo {author} {\bibfnamefont {Debdas}\ \bibnamefont {Dhabal}},
  \bibinfo {author} {\bibfnamefont {Andrew~Huy}\ \bibnamefont {Nguyen}},
  \bibinfo {author} {\bibfnamefont {Valeria}\ \bibnamefont {Molinero}}, \ and\
  \bibinfo {author} {\bibfnamefont {Charusita}\ \bibnamefont {Chakravarty}},\
  }\bibfield  {title} {\enquote {\bibinfo {title} {Triplet correlations
  dominate the transition from simple to tetrahedral liquids},}\ }\href@noop {}
  {\bibfield  {journal} {\bibinfo  {journal} {Phys. Rev. Lett.}\ }\textbf
  {\bibinfo {volume} {112}},\ \bibinfo {pages} {147801} (\bibinfo {year}
  {2014})}\BibitemShut {NoStop}%
\bibitem [{\citenamefont {Dhabal}\ \emph {et~al.}(2015)\citenamefont {Dhabal},
  \citenamefont {Nguyen}, \citenamefont {Singh}, \citenamefont {Khatua},
  \citenamefont {Molinero}, \citenamefont {Bandyopadhyay},\ and\ \citenamefont
  {Chakravarty}}]{dhabal2015excess}%
  \BibitemOpen
  \bibfield  {author} {\bibinfo {author} {\bibfnamefont {Debdas}\ \bibnamefont
  {Dhabal}}, \bibinfo {author} {\bibfnamefont {Andrew~Huy}\ \bibnamefont
  {Nguyen}}, \bibinfo {author} {\bibfnamefont {Murari}\ \bibnamefont {Singh}},
  \bibinfo {author} {\bibfnamefont {Prabir}\ \bibnamefont {Khatua}}, \bibinfo
  {author} {\bibfnamefont {Valeria}\ \bibnamefont {Molinero}}, \bibinfo
  {author} {\bibfnamefont {Sanjoy}\ \bibnamefont {Bandyopadhyay}}, \ and\
  \bibinfo {author} {\bibfnamefont {Charusita}\ \bibnamefont {Chakravarty}},\
  }\bibfield  {title} {\enquote {\bibinfo {title} {Excess entropy and
  crystallization in stillinger-weber and lennard-jones fluids},}\ }\href@noop
  {} {\bibfield  {journal} {\bibinfo  {journal} {J. Chem. Phys.}\ }\textbf
  {\bibinfo {volume} {143}},\ \bibinfo {pages} {164512} (\bibinfo {year}
  {2015})}\BibitemShut {NoStop}%
\bibitem [{\citenamefont {Angell}\ and\ \citenamefont
  {Kapko}(2016)}]{angell2016potential}%
  \BibitemOpen
  \bibfield  {author} {\bibinfo {author} {\bibfnamefont {C~Austen}\
  \bibnamefont {Angell}}\ and\ \bibinfo {author} {\bibfnamefont {Vitaliy}\
  \bibnamefont {Kapko}},\ }\bibfield  {title} {\enquote {\bibinfo {title}
  {Potential tuning in the s--w system.(i) bringing t c, 2 to ambient pressure,
  and (ii) colliding t c, 2 with the liquid--vapor spinodal},}\ }\href@noop {}
  {\bibfield  {journal} {\bibinfo  {journal} {J. Stat. Phys.}\ }\textbf
  {\bibinfo {volume} {2016}},\ \bibinfo {pages} {094004} (\bibinfo {year}
  {2016})}\BibitemShut {NoStop}%
\bibitem [{\citenamefont {Dhabal}\ \emph {et~al.}(2016)\citenamefont {Dhabal},
  \citenamefont {Chakravarty}, \citenamefont {Molinero},\ and\ \citenamefont
  {Kashyap}}]{dhabal}%
  \BibitemOpen
  \bibfield  {author} {\bibinfo {author} {\bibfnamefont {Debdas}\ \bibnamefont
  {Dhabal}}, \bibinfo {author} {\bibfnamefont {Charusita}\ \bibnamefont
  {Chakravarty}}, \bibinfo {author} {\bibfnamefont {Valeria}\ \bibnamefont
  {Molinero}}, \ and\ \bibinfo {author} {\bibfnamefont {Hemant~K.}\
  \bibnamefont {Kashyap}},\ }\bibfield  {title} {\enquote {\bibinfo {title}
  {Comparison of liquid-state anomalies in stillinger-weber models of water,
  silicon, and germanium},}\ }\href {\doibase 10.1063/1.4967939} {\bibfield
  {journal} {\bibinfo  {journal} {J. Chem. Phys.}\ }\textbf {\bibinfo {volume}
  {145}},\ \bibinfo {pages} {214502} (\bibinfo {year} {2016})}\BibitemShut
  {NoStop}%
\bibitem [{\citenamefont {Akahane}\ \emph {et~al.}(2016)\citenamefont
  {Akahane}, \citenamefont {Russo},\ and\ \citenamefont
  {Tanaka}}]{akahane_quadruple}%
  \BibitemOpen
  \bibfield  {author} {\bibinfo {author} {\bibfnamefont {Kenji}\ \bibnamefont
  {Akahane}}, \bibinfo {author} {\bibfnamefont {John}\ \bibnamefont {Russo}}, \
  and\ \bibinfo {author} {\bibfnamefont {Hajime}\ \bibnamefont {Tanaka}},\
  }\bibfield  {title} {\enquote {\bibinfo {title} {A possible four-phase
  coexistence in a single-component system},}\ }\href@noop {} {\bibfield
  {journal} {\bibinfo  {journal} {Nat. Commun.}\ }\textbf {\bibinfo {volume}
  {7}} (\bibinfo {year} {2016})}\BibitemShut {NoStop}%
\bibitem [{\citenamefont {Tanaka}(1998)}]{tanaka1998simple}%
  \BibitemOpen
  \bibfield  {author} {\bibinfo {author} {\bibfnamefont {Hajime}\ \bibnamefont
  {Tanaka}},\ }\bibfield  {title} {\enquote {\bibinfo {title} {Simple physical
  explanation of the unusual thermodynamic behavior of liquid water},}\
  }\href@noop {} {\bibfield  {journal} {\bibinfo  {journal} {Phys. Rev. Lett.}\
  }\textbf {\bibinfo {volume} {80}},\ \bibinfo {pages} {5750--5753} (\bibinfo
  {year} {1998})}\BibitemShut {NoStop}%
\bibitem [{\citenamefont {Tanaka}(2000{\natexlab{a}})}]{tanaka2000simple}%
  \BibitemOpen
  \bibfield  {author} {\bibinfo {author} {\bibfnamefont {Hajime}\ \bibnamefont
  {Tanaka}},\ }\bibfield  {title} {\enquote {\bibinfo {title} {Simple physical
  model of liquid water},}\ }\href@noop {} {\bibfield  {journal} {\bibinfo
  {journal} {J. Chem. Phys.}\ }\textbf {\bibinfo {volume} {112}},\ \bibinfo
  {pages} {799--809} (\bibinfo {year} {2000}{\natexlab{a}})}\BibitemShut
  {NoStop}%
\bibitem [{\citenamefont {Tanaka}(2000{\natexlab{b}})}]{Tanaka2000a}%
  \BibitemOpen
  \bibfield  {author} {\bibinfo {author} {\bibfnamefont {Hajime}\ \bibnamefont
  {Tanaka}},\ }\bibfield  {title} {\enquote {\bibinfo {title} {{Thermodynamic
  anomaly and polyamorphism of water}},}\ }\href {\doibase
  10.1209/epl/i2000-00276-4} {\bibfield  {journal} {\bibinfo  {journal}
  {Europhys. Lett.}\ }\textbf {\bibinfo {volume} {50}},\ \bibinfo {pages}
  {340--346} (\bibinfo {year} {2000}{\natexlab{b}})}\BibitemShut {NoStop}%
\bibitem [{\citenamefont {Tanaka}(2003)}]{Tanaka2003}%
  \BibitemOpen
  \bibfield  {author} {\bibinfo {author} {\bibfnamefont {Hajime}\ \bibnamefont
  {Tanaka}},\ }\bibfield  {title} {\enquote {\bibinfo {title} {{A new scenario
  of the apparent fragile-to-strong transition in tetrahedral liquids: water as
  an example}},}\ }\href {\doibase 10.1088/0953-8984/15/45/L03} {\bibfield
  {journal} {\bibinfo  {journal} {J. Phys.: Condens. Matter}\ }\textbf
  {\bibinfo {volume} {15}},\ \bibinfo {pages} {L703--L711} (\bibinfo {year}
  {2003})}\BibitemShut {NoStop}%
\bibitem [{\citenamefont {Russo}\ and\ \citenamefont
  {Tanaka}(2014)}]{russo2014understanding}%
  \BibitemOpen
  \bibfield  {author} {\bibinfo {author} {\bibfnamefont {John}\ \bibnamefont
  {Russo}}\ and\ \bibinfo {author} {\bibfnamefont {Hajime}\ \bibnamefont
  {Tanaka}},\ }\bibfield  {title} {\enquote {\bibinfo {title} {Understanding
  water anomalies with locally favoured structures},}\ }\href@noop {}
  {\bibfield  {journal} {\bibinfo  {journal} {Nat. Commun.}\ }\textbf {\bibinfo
  {volume} {5}} (\bibinfo {year} {2014})}\BibitemShut {NoStop}%
\bibitem [{\citenamefont {Holten}\ and\ \citenamefont
  {Anisimov}(2012)}]{holten2012entropy}%
  \BibitemOpen
  \bibfield  {author} {\bibinfo {author} {\bibfnamefont {V}~\bibnamefont
  {Holten}}\ and\ \bibinfo {author} {\bibfnamefont {MA}~\bibnamefont
  {Anisimov}},\ }\bibfield  {title} {\enquote {\bibinfo {title} {Entropy-driven
  liquid--liquid separation in supercooled water},}\ }\href@noop {} {\bibfield
  {journal} {\bibinfo  {journal} {Sci. Rep.}\ }\textbf {\bibinfo {volume} {2}}
  (\bibinfo {year} {2012})}\BibitemShut {NoStop}%
\bibitem [{\citenamefont {Holten}\ \emph
  {et~al.}(2013{\natexlab{b}})\citenamefont {Holten}, \citenamefont {Limmer},
  \citenamefont {Molinero},\ and\ \citenamefont {Anisimov}}]{holten2013nature}%
  \BibitemOpen
  \bibfield  {author} {\bibinfo {author} {\bibfnamefont {Vincent}\ \bibnamefont
  {Holten}}, \bibinfo {author} {\bibfnamefont {David~T}\ \bibnamefont
  {Limmer}}, \bibinfo {author} {\bibfnamefont {Valeria}\ \bibnamefont
  {Molinero}}, \ and\ \bibinfo {author} {\bibfnamefont {Mikhail~A}\
  \bibnamefont {Anisimov}},\ }\bibfield  {title} {\enquote {\bibinfo {title}
  {Nature of the anomalies in the supercooled liquid state of the mw model of
  water},}\ }\href@noop {} {\bibfield  {journal} {\bibinfo  {journal} {J. Chem.
  Phys.}\ }\textbf {\bibinfo {volume} {138}},\ \bibinfo {pages} {174501}
  (\bibinfo {year} {2013}{\natexlab{b}})}\BibitemShut {NoStop}%
\bibitem [{\citenamefont {Holten}\ \emph {et~al.}(2014)\citenamefont {Holten},
  \citenamefont {Palmer}, \citenamefont {Poole}, \citenamefont {Debenedetti},\
  and\ \citenamefont {Anisimov}}]{holten2014two}%
  \BibitemOpen
  \bibfield  {author} {\bibinfo {author} {\bibfnamefont {Vincent}\ \bibnamefont
  {Holten}}, \bibinfo {author} {\bibfnamefont {Jeremy~C}\ \bibnamefont
  {Palmer}}, \bibinfo {author} {\bibfnamefont {Peter~H}\ \bibnamefont {Poole}},
  \bibinfo {author} {\bibfnamefont {Pablo~G}\ \bibnamefont {Debenedetti}}, \
  and\ \bibinfo {author} {\bibfnamefont {Mikhail~A}\ \bibnamefont {Anisimov}},\
  }\bibfield  {title} {\enquote {\bibinfo {title} {Two-state thermodynamics of
  the st2 model for supercooled water},}\ }\href@noop {} {\bibfield  {journal}
  {\bibinfo  {journal} {J. Chem. Phys.}\ }\textbf {\bibinfo {volume} {140}},\
  \bibinfo {pages} {104502} (\bibinfo {year} {2014})}\BibitemShut {NoStop}%
\bibitem [{\citenamefont {R{\"o}ntgen}(1892)}]{rontgen1892ueber}%
  \BibitemOpen
  \bibfield  {author} {\bibinfo {author} {\bibfnamefont {Wilhelm~Conrad}\
  \bibnamefont {R{\"o}ntgen}},\ }\bibfield  {title} {\enquote {\bibinfo {title}
  {Ueber die constitution des fl{\"u}ssigen wassers},}\ }\href@noop {}
  {\bibfield  {journal} {\bibinfo  {journal} {Ann. Phys.}\ }\textbf {\bibinfo
  {volume} {281}},\ \bibinfo {pages} {91--97} (\bibinfo {year}
  {1892})}\BibitemShut {NoStop}%
\bibitem [{\citenamefont {Angell}(1971)}]{Angell_w}%
  \BibitemOpen
  \bibfield  {author} {\bibinfo {author} {\bibfnamefont {C.~A.}\ \bibnamefont
  {Angell}},\ }\bibfield  {title} {\enquote {\bibinfo {title} {Two-state
  thermodynamics and transport properties for water from "bond lattice"
  model},}\ }\href@noop {} {\bibfield  {journal} {\bibinfo  {journal} {J. Phys.
  Chem.}\ }\textbf {\bibinfo {volume} {75}},\ \bibinfo {pages} {3698--3705}
  (\bibinfo {year} {1971})}\BibitemShut {NoStop}%
\bibitem [{\citenamefont {Cho}\ \emph {et~al.}(1996)\citenamefont {Cho},
  \citenamefont {Singh},\ and\ \citenamefont {Robinson}}]{robinson}%
  \BibitemOpen
  \bibfield  {author} {\bibinfo {author} {\bibfnamefont {Chul~Hee}\
  \bibnamefont {Cho}}, \bibinfo {author} {\bibfnamefont {Surjit}\ \bibnamefont
  {Singh}}, \ and\ \bibinfo {author} {\bibfnamefont {G.~Wilse}\ \bibnamefont
  {Robinson}},\ }\bibfield  {title} {\enquote {\bibinfo {title} {An explanation
  of the density maximum in water},}\ }\href {\doibase
  10.1103/PhysRevLett.76.1651} {\bibfield  {journal} {\bibinfo  {journal}
  {Phys. Rev. Lett.}\ }\textbf {\bibinfo {volume} {76}},\ \bibinfo {pages}
  {1651--1654} (\bibinfo {year} {1996})}\BibitemShut {NoStop}%
\bibitem [{\citenamefont {Ponyatovsky}\ \emph {et~al.}(1998)\citenamefont
  {Ponyatovsky}, \citenamefont {Sinitsyn},\ and\ \citenamefont
  {Pozdnyakova}}]{ponyatovsky}%
  \BibitemOpen
  \bibfield  {author} {\bibinfo {author} {\bibfnamefont {E.~G.}\ \bibnamefont
  {Ponyatovsky}}, \bibinfo {author} {\bibfnamefont {V.~V.}\ \bibnamefont
  {Sinitsyn}}, \ and\ \bibinfo {author} {\bibfnamefont {T.~A.}\ \bibnamefont
  {Pozdnyakova}},\ }\bibfield  {title} {\enquote {\bibinfo {title} {The
  metastable $t$-$p$ phase diagram and anomalous thermodynamic properties of
  supercooled water},}\ }\href@noop {} {\bibfield  {journal} {\bibinfo
  {journal} {J. Chem. Phys.}\ }\textbf {\bibinfo {volume} {109}},\ \bibinfo
  {pages} {2413--2422} (\bibinfo {year} {1998})}\BibitemShut {NoStop}%
\bibitem [{\citenamefont {N{\'e}methy}\ and\ \citenamefont
  {Scheraga}(1962)}]{nemethy1962structure}%
  \BibitemOpen
  \bibfield  {author} {\bibinfo {author} {\bibfnamefont {George}\ \bibnamefont
  {N{\'e}methy}}\ and\ \bibinfo {author} {\bibfnamefont {Harold~A.}\
  \bibnamefont {Scheraga}},\ }\bibfield  {title} {\enquote {\bibinfo {title}
  {Structure of water and hydrophobic bonding in proteins. i. a model for the
  thermodynamic properties of liquid water},}\ }\href@noop {} {\bibfield
  {journal} {\bibinfo  {journal} {J. Chem. Phys.}\ }\textbf {\bibinfo {volume}
  {36}},\ \bibinfo {pages} {3382--3400} (\bibinfo {year} {1962})}\BibitemShut
  {NoStop}%
\bibitem [{\citenamefont {Anisimov}\ \emph {et~al.}(2018)\citenamefont
  {Anisimov}, \citenamefont {Du{\v{s}}ka}, \citenamefont {Caupin},
  \citenamefont {Amrhein}, \citenamefont {Rosenbaum},\ and\ \citenamefont
  {Sadus}}]{anisimov2018thermodynamics}%
  \BibitemOpen
  \bibfield  {author} {\bibinfo {author} {\bibfnamefont {Mikhail~A}\
  \bibnamefont {Anisimov}}, \bibinfo {author} {\bibfnamefont {Michal}\
  \bibnamefont {Du{\v{s}}ka}}, \bibinfo {author} {\bibfnamefont
  {Fr{\'e}d{\'e}ric}\ \bibnamefont {Caupin}}, \bibinfo {author} {\bibfnamefont
  {Lauren~E}\ \bibnamefont {Amrhein}}, \bibinfo {author} {\bibfnamefont
  {Amanda}\ \bibnamefont {Rosenbaum}}, \ and\ \bibinfo {author} {\bibfnamefont
  {Richard~J}\ \bibnamefont {Sadus}},\ }\bibfield  {title} {\enquote {\bibinfo
  {title} {Thermodynamics of fluid polyamorphism},}\ }\href@noop {} {\bibfield
  {journal} {\bibinfo  {journal} {Physical Review X}\ }\textbf {\bibinfo
  {volume} {8}},\ \bibinfo {pages} {011004} (\bibinfo {year}
  {2018})}\BibitemShut {NoStop}%
\bibitem [{\citenamefont {Tanaka}(2012{\natexlab{a}})}]{tanaka_review}%
  \BibitemOpen
  \bibfield  {author} {\bibinfo {author} {\bibfnamefont {H.}~\bibnamefont
  {Tanaka}},\ }\bibfield  {title} {\enquote {\bibinfo {title} {Bond
  orientational order in liquids: Towards a unified description of water-like
  anomalies, liquid-liquid transition, glass transition, and
  crystallization},}\ }\href@noop {} {\bibfield  {journal} {\bibinfo  {journal}
  {Eur. Phys. J E}\ }\textbf {\bibinfo {volume} {35}},\ \bibinfo {pages} {113}
  (\bibinfo {year} {2012}{\natexlab{a}})}\BibitemShut {NoStop}%
\bibitem [{\citenamefont {Tanaka}(1999)}]{tanaka1999two}%
  \BibitemOpen
  \bibfield  {author} {\bibinfo {author} {\bibfnamefont {Hajime}\ \bibnamefont
  {Tanaka}},\ }\bibfield  {title} {\enquote {\bibinfo {title}
  {Two-order-parameter description of liquids: critical phenomena and phase
  separation of supercooled liquids},}\ }\href@noop {} {\bibfield  {journal}
  {\bibinfo  {journal} {J. Phys.: Condens. Matter}\ }\textbf {\bibinfo {volume}
  {11}},\ \bibinfo {pages} {L159} (\bibinfo {year} {1999})}\BibitemShut
  {NoStop}%
\bibitem [{\citenamefont {Errington}\ and\ \citenamefont
  {Debenedetti}(2001)}]{errington2001relationship}%
  \BibitemOpen
  \bibfield  {author} {\bibinfo {author} {\bibfnamefont {Jeffrey~R}\
  \bibnamefont {Errington}}\ and\ \bibinfo {author} {\bibfnamefont {Pablo~G}\
  \bibnamefont {Debenedetti}},\ }\bibfield  {title} {\enquote {\bibinfo {title}
  {Relationship between structural order and the anomalies of liquid water},}\
  }\href@noop {} {\bibfield  {journal} {\bibinfo  {journal} {Nature}\ }\textbf
  {\bibinfo {volume} {409}},\ \bibinfo {pages} {318--321} (\bibinfo {year}
  {2001})}\BibitemShut {NoStop}%
\bibitem [{\citenamefont {Cuthbertson}\ and\ \citenamefont
  {Poole}(2011)}]{cuthbertson2011mixturelike}%
  \BibitemOpen
  \bibfield  {author} {\bibinfo {author} {\bibfnamefont {Megan~J}\ \bibnamefont
  {Cuthbertson}}\ and\ \bibinfo {author} {\bibfnamefont {Peter~H}\ \bibnamefont
  {Poole}},\ }\bibfield  {title} {\enquote {\bibinfo {title} {Mixturelike
  behavior near a liquid-liquid phase transition in simulations of supercooled
  water},}\ }\href@noop {} {\bibfield  {journal} {\bibinfo  {journal} {Phys.
  Rev. Lett.}\ }\textbf {\bibinfo {volume} {106}},\ \bibinfo {pages} {115706}
  (\bibinfo {year} {2011})}\BibitemShut {NoStop}%
\bibitem [{\citenamefont {Wikfeldt}\ \emph {et~al.}(2011)\citenamefont
  {Wikfeldt}, \citenamefont {Nilsson},\ and\ \citenamefont
  {Pettersson}}]{wikfeldt2011spatially}%
  \BibitemOpen
  \bibfield  {author} {\bibinfo {author} {\bibfnamefont {KT}~\bibnamefont
  {Wikfeldt}}, \bibinfo {author} {\bibfnamefont {Anders}\ \bibnamefont
  {Nilsson}}, \ and\ \bibinfo {author} {\bibfnamefont {Lars~GM}\ \bibnamefont
  {Pettersson}},\ }\bibfield  {title} {\enquote {\bibinfo {title} {Spatially
  inhomogeneous bimodal inherent structure of simulated liquid water},}\
  }\href@noop {} {\bibfield  {journal} {\bibinfo  {journal} {Phys. Chem. Chem.
  Phys.}\ }\textbf {\bibinfo {volume} {13}},\ \bibinfo {pages} {19918--19924}
  (\bibinfo {year} {2011})}\BibitemShut {NoStop}%
\bibitem [{\citenamefont {Tanaka}(2012{\natexlab{b}})}]{tanaka2012bond}%
  \BibitemOpen
  \bibfield  {author} {\bibinfo {author} {\bibfnamefont {Hajime}\ \bibnamefont
  {Tanaka}},\ }\bibfield  {title} {\enquote {\bibinfo {title} {Bond
  orientational order in liquids: Towards a unified description of water-like
  anomalies, liquid-liquid transition, glass transition, and
  crystallization},}\ }\href@noop {} {\bibfield  {journal} {\bibinfo  {journal}
  {Eur. Phys. J. E}\ }\textbf {\bibinfo {volume} {35}},\ \bibinfo {pages}
  {1--84} (\bibinfo {year} {2012}{\natexlab{b}})}\BibitemShut {NoStop}%
\bibitem [{\citenamefont {Xu}\ \emph {et~al.}(2005)\citenamefont {Xu},
  \citenamefont {Kumar}, \citenamefont {Buldyrev}, \citenamefont {Chen},
  \citenamefont {Poole}, \citenamefont {Sciortino},\ and\ \citenamefont
  {Stanley}}]{xu2005relation}%
  \BibitemOpen
  \bibfield  {author} {\bibinfo {author} {\bibfnamefont {Limei}\ \bibnamefont
  {Xu}}, \bibinfo {author} {\bibfnamefont {Pradeep}\ \bibnamefont {Kumar}},
  \bibinfo {author} {\bibfnamefont {Sergey~V}\ \bibnamefont {Buldyrev}},
  \bibinfo {author} {\bibfnamefont {S-H}\ \bibnamefont {Chen}}, \bibinfo
  {author} {\bibfnamefont {Peter~H}\ \bibnamefont {Poole}}, \bibinfo {author}
  {\bibfnamefont {Francesco}\ \bibnamefont {Sciortino}}, \ and\ \bibinfo
  {author} {\bibfnamefont {H~Eugene}\ \bibnamefont {Stanley}},\ }\bibfield
  {title} {\enquote {\bibinfo {title} {Relation between the widom line and the
  dynamic crossover in systems with a liquid--liquid phase transition},}\
  }\href@noop {} {\bibfield  {journal} {\bibinfo  {journal} {Proc. Natl. Acad.
  Sci. USA}\ }\textbf {\bibinfo {volume} {102}},\ \bibinfo {pages}
  {16558--16562} (\bibinfo {year} {2005})}\BibitemShut {NoStop}%
\bibitem [{\citenamefont {Cerveny}\ \emph {et~al.}(2016)\citenamefont
  {Cerveny}, \citenamefont {Mallamace}, \citenamefont {Swenson}, \citenamefont
  {Vogel},\ and\ \citenamefont {Xu}}]{cerveny2016confined}%
  \BibitemOpen
  \bibfield  {author} {\bibinfo {author} {\bibfnamefont {Silvina}\ \bibnamefont
  {Cerveny}}, \bibinfo {author} {\bibfnamefont {Francesco}\ \bibnamefont
  {Mallamace}}, \bibinfo {author} {\bibfnamefont {Jan}\ \bibnamefont
  {Swenson}}, \bibinfo {author} {\bibfnamefont {Michael}\ \bibnamefont
  {Vogel}}, \ and\ \bibinfo {author} {\bibfnamefont {Limei}\ \bibnamefont
  {Xu}},\ }\bibfield  {title} {\enquote {\bibinfo {title} {Confined water as
  model of supercooled water},}\ }\href@noop {} {\bibfield  {journal} {\bibinfo
   {journal} {Chem. Rev.}\ }\textbf {\bibinfo {volume} {116}},\ \bibinfo
  {pages} {7608--7625} (\bibinfo {year} {2016})}\BibitemShut {NoStop}%
\bibitem [{\citenamefont {Faraone}\ \emph {et~al.}(2004)\citenamefont
  {Faraone}, \citenamefont {Liu}, \citenamefont {Mou}, \citenamefont {Yen},\
  and\ \citenamefont {Chen}}]{faraone2004fragile}%
  \BibitemOpen
  \bibfield  {author} {\bibinfo {author} {\bibfnamefont {A}~\bibnamefont
  {Faraone}}, \bibinfo {author} {\bibfnamefont {Li}~\bibnamefont {Liu}},
  \bibinfo {author} {\bibfnamefont {C-Y}\ \bibnamefont {Mou}}, \bibinfo
  {author} {\bibfnamefont {C-W}\ \bibnamefont {Yen}}, \ and\ \bibinfo {author}
  {\bibfnamefont {S-H}\ \bibnamefont {Chen}},\ }\bibfield  {title} {\enquote
  {\bibinfo {title} {Fragile-to-strong liquid transition in deeply supercooled
  confined water},}\ }\href@noop {} {\bibfield  {journal} {\bibinfo  {journal}
  {J. Chem. Phys.}\ }\textbf {\bibinfo {volume} {121}},\ \bibinfo {pages}
  {10843--10846} (\bibinfo {year} {2004})}\BibitemShut {NoStop}%
\bibitem [{\citenamefont {Zhang}\ \emph {et~al.}(2009)\citenamefont {Zhang},
  \citenamefont {Lagi}, \citenamefont {Fratini}, \citenamefont {Baglioni},
  \citenamefont {Mamontov},\ and\ \citenamefont {Chen}}]{zhang2009dynamic}%
  \BibitemOpen
  \bibfield  {author} {\bibinfo {author} {\bibfnamefont {Yang}\ \bibnamefont
  {Zhang}}, \bibinfo {author} {\bibfnamefont {Marco}\ \bibnamefont {Lagi}},
  \bibinfo {author} {\bibfnamefont {Emiliano}\ \bibnamefont {Fratini}},
  \bibinfo {author} {\bibfnamefont {Piero}\ \bibnamefont {Baglioni}}, \bibinfo
  {author} {\bibfnamefont {Eugene}\ \bibnamefont {Mamontov}}, \ and\ \bibinfo
  {author} {\bibfnamefont {Sow-Hsin}\ \bibnamefont {Chen}},\ }\bibfield
  {title} {\enquote {\bibinfo {title} {Dynamic susceptibility of supercooled
  water and its relation to the dynamic crossover phenomenon},}\ }\href@noop {}
  {\bibfield  {journal} {\bibinfo  {journal} {Phys. Rev. E}\ }\textbf {\bibinfo
  {volume} {79}},\ \bibinfo {pages} {040201} (\bibinfo {year}
  {2009})}\BibitemShut {NoStop}%
\bibitem [{\citenamefont {Gallo}\ \emph {et~al.}(2010)\citenamefont {Gallo},
  \citenamefont {Rovere},\ and\ \citenamefont {Chen}}]{gallo2010dynamic}%
  \BibitemOpen
  \bibfield  {author} {\bibinfo {author} {\bibfnamefont {P}~\bibnamefont
  {Gallo}}, \bibinfo {author} {\bibfnamefont {M}~\bibnamefont {Rovere}}, \ and\
  \bibinfo {author} {\bibfnamefont {S-H}\ \bibnamefont {Chen}},\ }\bibfield
  {title} {\enquote {\bibinfo {title} {Dynamic crossover in supercooled
  confined water: understanding bulk properties through confinement},}\
  }\href@noop {} {\bibfield  {journal} {\bibinfo  {journal} {J. Phys. Chem.
  Lett.}\ }\textbf {\bibinfo {volume} {1}},\ \bibinfo {pages} {729--733}
  (\bibinfo {year} {2010})}\BibitemShut {NoStop}%
\bibitem [{\citenamefont {Wang}\ \emph {et~al.}(2015)\citenamefont {Wang},
  \citenamefont {Le}, \citenamefont {Ito}, \citenamefont {Le{\~a}o},
  \citenamefont {Tyagi},\ and\ \citenamefont {Chen}}]{wang2015dynamic}%
  \BibitemOpen
  \bibfield  {author} {\bibinfo {author} {\bibfnamefont {Zhe}\ \bibnamefont
  {Wang}}, \bibinfo {author} {\bibfnamefont {Peisi}\ \bibnamefont {Le}},
  \bibinfo {author} {\bibfnamefont {Kanae}\ \bibnamefont {Ito}}, \bibinfo
  {author} {\bibfnamefont {Juscelino~B}\ \bibnamefont {Le{\~a}o}}, \bibinfo
  {author} {\bibfnamefont {Madhusudan}\ \bibnamefont {Tyagi}}, \ and\ \bibinfo
  {author} {\bibfnamefont {Sow-Hsin}\ \bibnamefont {Chen}},\ }\bibfield
  {title} {\enquote {\bibinfo {title} {Dynamic crossover in deeply cooled water
  confined in mcm-41 at 4 kbar and its relation to the liquid-liquid transition
  hypothesis},}\ }\href@noop {} {\bibfield  {journal} {\bibinfo  {journal} {J.
  Chem. Phys.}\ }\textbf {\bibinfo {volume} {143}},\ \bibinfo {pages} {114508}
  (\bibinfo {year} {2015})}\BibitemShut {NoStop}%
\bibitem [{\citenamefont {Shi}\ \emph {et~al.}(2017)\citenamefont {Shi},
  \citenamefont {Russo},\ and\ \citenamefont {Tanaka}}]{shi2017}%
  \BibitemOpen
  \bibfield  {author} {\bibinfo {author} {\bibfnamefont {Rui}\ \bibnamefont
  {Shi}}, \bibinfo {author} {\bibfnamefont {John}\ \bibnamefont {Russo}}, \
  and\ \bibinfo {author} {\bibfnamefont {Hajime}\ \bibnamefont {Tanaka}},\
  }\bibfield  {title} {\enquote {\bibinfo {title} {Common microscopic
  structural origin for water's thermodynamic and dynamic anomalies},}\
  }\href@noop {} {\bibfield  {journal} {\bibinfo  {journal} {in preparation}\ }
  (\bibinfo {year} {2017})}\BibitemShut {NoStop}%
\bibitem [{\citenamefont {Singh}\ \emph {et~al.}(2017)\citenamefont {Singh},
  \citenamefont {Issenmann},\ and\ \citenamefont {Caupin}}]{singh2017pressure}%
  \BibitemOpen
  \bibfield  {author} {\bibinfo {author} {\bibfnamefont {Lokendra~P}\
  \bibnamefont {Singh}}, \bibinfo {author} {\bibfnamefont {Bruno}\ \bibnamefont
  {Issenmann}}, \ and\ \bibinfo {author} {\bibfnamefont {Fr{\'e}d{\'e}ric}\
  \bibnamefont {Caupin}},\ }\bibfield  {title} {\enquote {\bibinfo {title}
  {Pressure dependence of viscosity in supercooled water and a unified approach
  for thermodynamic and dynamic anomalies of water},}\ }\href@noop {}
  {\bibfield  {journal} {\bibinfo  {journal} {Proceedings of the National
  Academy of Sciences}\ ,\ \bibinfo {pages} {201619501}} (\bibinfo {year}
  {2017})}\BibitemShut {NoStop}%
\bibitem [{\citenamefont {Florusse}\ \emph {et~al.}(2004)\citenamefont
  {Florusse}, \citenamefont {Peters}, \citenamefont {Schoonman}, \citenamefont
  {Hester}, \citenamefont {Koh}, \citenamefont {Dec}, \citenamefont {Marsh},\
  and\ \citenamefont {Sloan}}]{florusse2004stable}%
  \BibitemOpen
  \bibfield  {author} {\bibinfo {author} {\bibfnamefont {Louw~J}\ \bibnamefont
  {Florusse}}, \bibinfo {author} {\bibfnamefont {Cor~J}\ \bibnamefont
  {Peters}}, \bibinfo {author} {\bibfnamefont {Joop}\ \bibnamefont
  {Schoonman}}, \bibinfo {author} {\bibfnamefont {Keith~C}\ \bibnamefont
  {Hester}}, \bibinfo {author} {\bibfnamefont {Carolyn~A}\ \bibnamefont {Koh}},
  \bibinfo {author} {\bibfnamefont {Steven~F}\ \bibnamefont {Dec}}, \bibinfo
  {author} {\bibfnamefont {Kenneth~N}\ \bibnamefont {Marsh}}, \ and\ \bibinfo
  {author} {\bibfnamefont {E~Dendy}\ \bibnamefont {Sloan}},\ }\bibfield
  {title} {\enquote {\bibinfo {title} {Stable low-pressure hydrogen clusters
  stored in a binary clathrate hydrate},}\ }\href@noop {} {\bibfield  {journal}
  {\bibinfo  {journal} {Science}\ }\textbf {\bibinfo {volume} {306}},\ \bibinfo
  {pages} {469--471} (\bibinfo {year} {2004})}\BibitemShut {NoStop}%
\bibitem [{\citenamefont {Lee}\ \emph {et~al.}(2005)\citenamefont {Lee},
  \citenamefont {Lee}, \citenamefont {Park}, \citenamefont {Seo}, \citenamefont
  {Zeng}, \citenamefont {Moudrakovski}, \citenamefont {Ratcliffe},
  \citenamefont {Ripmeester} \emph {et~al.}}]{lee2005tuning}%
  \BibitemOpen
  \bibfield  {author} {\bibinfo {author} {\bibfnamefont {Huen}\ \bibnamefont
  {Lee}}, \bibinfo {author} {\bibfnamefont {Jong-won}\ \bibnamefont {Lee}},
  \bibinfo {author} {\bibfnamefont {Jeasung}\ \bibnamefont {Park}}, \bibinfo
  {author} {\bibfnamefont {Yu-Taek}\ \bibnamefont {Seo}}, \bibinfo {author}
  {\bibfnamefont {Huang}\ \bibnamefont {Zeng}}, \bibinfo {author}
  {\bibfnamefont {Igor~L}\ \bibnamefont {Moudrakovski}}, \bibinfo {author}
  {\bibfnamefont {Christopher~I}\ \bibnamefont {Ratcliffe}}, \bibinfo {author}
  {\bibfnamefont {John~A}\ \bibnamefont {Ripmeester}},  \emph {et~al.},\
  }\bibfield  {title} {\enquote {\bibinfo {title} {Tuning clathrate hydrates
  for hydrogen storage},}\ }\href@noop {} {\bibfield  {journal} {\bibinfo
  {journal} {Nature}\ }\textbf {\bibinfo {volume} {434}},\ \bibinfo {pages}
  {743--746} (\bibinfo {year} {2005})}\BibitemShut {NoStop}%
\bibitem [{\citenamefont {Chatti}\ \emph {et~al.}(2005)\citenamefont {Chatti},
  \citenamefont {Delahaye}, \citenamefont {Fournaison},\ and\ \citenamefont
  {Petitet}}]{chatti2005benefits}%
  \BibitemOpen
  \bibfield  {author} {\bibinfo {author} {\bibfnamefont {Imen}\ \bibnamefont
  {Chatti}}, \bibinfo {author} {\bibfnamefont {Anthony}\ \bibnamefont
  {Delahaye}}, \bibinfo {author} {\bibfnamefont {Laurence}\ \bibnamefont
  {Fournaison}}, \ and\ \bibinfo {author} {\bibfnamefont {Jean-Pierre}\
  \bibnamefont {Petitet}},\ }\bibfield  {title} {\enquote {\bibinfo {title}
  {Benefits and drawbacks of clathrate hydrates: a review of their areas of
  interest},}\ }\href@noop {} {\bibfield  {journal} {\bibinfo  {journal}
  {Energy Conversion and Management}\ }\textbf {\bibinfo {volume} {46}},\
  \bibinfo {pages} {1333--1343} (\bibinfo {year} {2005})}\BibitemShut {NoStop}%
\bibitem [{\citenamefont {Struzhkin}\ \emph {et~al.}(2007)\citenamefont
  {Struzhkin}, \citenamefont {Militzer}, \citenamefont {Mao}, \citenamefont
  {Mao},\ and\ \citenamefont {Hemley}}]{struzhkin2007hydrogen}%
  \BibitemOpen
  \bibfield  {author} {\bibinfo {author} {\bibfnamefont {Viktor~V}\
  \bibnamefont {Struzhkin}}, \bibinfo {author} {\bibfnamefont {Burkhard}\
  \bibnamefont {Militzer}}, \bibinfo {author} {\bibfnamefont {Wendy~L}\
  \bibnamefont {Mao}}, \bibinfo {author} {\bibfnamefont {Ho-kwang}\
  \bibnamefont {Mao}}, \ and\ \bibinfo {author} {\bibfnamefont {Russell~J}\
  \bibnamefont {Hemley}},\ }\bibfield  {title} {\enquote {\bibinfo {title}
  {Hydrogen storage in molecular clathrates},}\ }\href@noop {} {\bibfield
  {journal} {\bibinfo  {journal} {Chem. Rev.}\ }\textbf {\bibinfo {volume}
  {107}},\ \bibinfo {pages} {4133--4151} (\bibinfo {year} {2007})}\BibitemShut
  {NoStop}%
\bibitem [{\citenamefont {Azouzi}\ \emph {et~al.}(2013)\citenamefont {Azouzi},
  \citenamefont {Ramboz}, \citenamefont {Lenain},\ and\ \citenamefont
  {Caupin}}]{azouzi2013coherent}%
  \BibitemOpen
  \bibfield  {author} {\bibinfo {author} {\bibfnamefont {Mouna El~Mekki}\
  \bibnamefont {Azouzi}}, \bibinfo {author} {\bibfnamefont {Claire}\
  \bibnamefont {Ramboz}}, \bibinfo {author} {\bibfnamefont
  {Jean-Fran{\c{c}}ois}\ \bibnamefont {Lenain}}, \ and\ \bibinfo {author}
  {\bibfnamefont {Fr{\'e}d{\'e}ric}\ \bibnamefont {Caupin}},\ }\bibfield
  {title} {\enquote {\bibinfo {title} {A coherent picture of water at extreme
  negative pressure},}\ }\href@noop {} {\bibfield  {journal} {\bibinfo
  {journal} {Nature Phys.}\ }\textbf {\bibinfo {volume} {9}},\ \bibinfo {pages}
  {38--41} (\bibinfo {year} {2013})}\BibitemShut {NoStop}%
\bibitem [{\citenamefont {Pallares}\ \emph {et~al.}(2014)\citenamefont
  {Pallares}, \citenamefont {Azouzi}, \citenamefont {Gonz{\'a}lez},
  \citenamefont {Aragones}, \citenamefont {Abascal}, \citenamefont
  {Valeriani},\ and\ \citenamefont {Caupin}}]{pallares2014anomalies}%
  \BibitemOpen
  \bibfield  {author} {\bibinfo {author} {\bibfnamefont {Ga{\"e}l}\
  \bibnamefont {Pallares}}, \bibinfo {author} {\bibfnamefont {Mouna El~Mekki}\
  \bibnamefont {Azouzi}}, \bibinfo {author} {\bibfnamefont {Miguel~A.}\
  \bibnamefont {Gonz{\'a}lez}}, \bibinfo {author} {\bibfnamefont {Juan~L.}\
  \bibnamefont {Aragones}}, \bibinfo {author} {\bibfnamefont {Jos{\'e} L.~F.}\
  \bibnamefont {Abascal}}, \bibinfo {author} {\bibfnamefont {Chantal}\
  \bibnamefont {Valeriani}}, \ and\ \bibinfo {author} {\bibfnamefont
  {Fr{\'e}d{\'e}ric}\ \bibnamefont {Caupin}},\ }\bibfield  {title} {\enquote
  {\bibinfo {title} {Anomalies in bulk supercooled water at negative
  pressure},}\ }\href@noop {} {\bibfield  {journal} {\bibinfo  {journal} {Proc.
  Natl. Acad. Sci. USA}\ }\textbf {\bibinfo {volume} {111}},\ \bibinfo {pages}
  {7936--7941} (\bibinfo {year} {2014})}\BibitemShut {NoStop}%
\bibitem [{\citenamefont {Gonz{\'a}lez}\ \emph {et~al.}(2016)\citenamefont
  {Gonz{\'a}lez}, \citenamefont {Valeriani}, \citenamefont {Caupin},\ and\
  \citenamefont {Abascal}}]{gonzalez2016comprehensive}%
  \BibitemOpen
  \bibfield  {author} {\bibinfo {author} {\bibfnamefont {Miguel~A}\
  \bibnamefont {Gonz{\'a}lez}}, \bibinfo {author} {\bibfnamefont {Chantal}\
  \bibnamefont {Valeriani}}, \bibinfo {author} {\bibfnamefont
  {Fr{\'e}d{\'e}ric}\ \bibnamefont {Caupin}}, \ and\ \bibinfo {author}
  {\bibfnamefont {Jos{\'e}~LF}\ \bibnamefont {Abascal}},\ }\bibfield  {title}
  {\enquote {\bibinfo {title} {A comprehensive scenario of the thermodynamic
  anomalies of water using the tip4p/2005 model},}\ }\href@noop {} {\bibfield
  {journal} {\bibinfo  {journal} {J. Chem. Phys.}\ }\textbf {\bibinfo {volume}
  {145}},\ \bibinfo {pages} {054505} (\bibinfo {year} {2016})}\BibitemShut
  {NoStop}%
\bibitem [{\citenamefont {Holten}\ \emph {et~al.}(2017)\citenamefont {Holten},
  \citenamefont {Qiu}, \citenamefont {Guillerm}, \citenamefont {Wilke},
  \citenamefont {Ri{\v{c}}ka}, \citenamefont {Frenz},\ and\ \citenamefont
  {Caupin}}]{holten2017compressibility}%
  \BibitemOpen
  \bibfield  {author} {\bibinfo {author} {\bibfnamefont {Vincent}\ \bibnamefont
  {Holten}}, \bibinfo {author} {\bibfnamefont {Chen}\ \bibnamefont {Qiu}},
  \bibinfo {author} {\bibfnamefont {Emmanuel}\ \bibnamefont {Guillerm}},
  \bibinfo {author} {\bibfnamefont {Max}\ \bibnamefont {Wilke}}, \bibinfo
  {author} {\bibfnamefont {Jaroslav}\ \bibnamefont {Ri{\v{c}}ka}}, \bibinfo
  {author} {\bibfnamefont {Martin}\ \bibnamefont {Frenz}}, \ and\ \bibinfo
  {author} {\bibfnamefont {Fr{\'e}d{\'e}ric}\ \bibnamefont {Caupin}},\
  }\bibfield  {title} {\enquote {\bibinfo {title} {Compressibility anomalies in
  stretched water and their interplay with density anomalies},}\ }\href@noop {}
  {\bibfield  {journal} {\bibinfo  {journal} {arXiv preprint arXiv:1708.00063}\
  } (\bibinfo {year} {2017})}\BibitemShut {NoStop}%
\bibitem [{\citenamefont {Speedy}(1982)}]{speedy1982stability}%
  \BibitemOpen
  \bibfield  {author} {\bibinfo {author} {\bibfnamefont {Robin~J}\ \bibnamefont
  {Speedy}},\ }\bibfield  {title} {\enquote {\bibinfo {title} {Stability-limit
  conjecture. an interpretation of the properties of water},}\ }\href@noop {}
  {\bibfield  {journal} {\bibinfo  {journal} {J. Chem. Phys.}\ }\textbf
  {\bibinfo {volume} {86}},\ \bibinfo {pages} {982--991} (\bibinfo {year}
  {1982})}\BibitemShut {NoStop}%
\bibitem [{\citenamefont {Romano}\ \emph {et~al.}(2014)\citenamefont {Romano},
  \citenamefont {Russo},\ and\ \citenamefont {Tanaka}}]{romano2014novel}%
  \BibitemOpen
  \bibfield  {author} {\bibinfo {author} {\bibfnamefont {Flavio}\ \bibnamefont
  {Romano}}, \bibinfo {author} {\bibfnamefont {John}\ \bibnamefont {Russo}}, \
  and\ \bibinfo {author} {\bibfnamefont {Hajime}\ \bibnamefont {Tanaka}},\
  }\bibfield  {title} {\enquote {\bibinfo {title} {Novel stable crystalline
  phase for the stillinger-weber potential},}\ }\href@noop {} {\bibfield
  {journal} {\bibinfo  {journal} {Phys. Rev. B}\ }\textbf {\bibinfo {volume}
  {90}},\ \bibinfo {pages} {014204} (\bibinfo {year} {2014})}\BibitemShut
  {NoStop}%
\bibitem [{\citenamefont {Kennett}\ \emph {et~al.}(2005)\citenamefont
  {Kennett}, \citenamefont {Chamon},\ and\ \citenamefont
  {Cugliandolo}}]{kennett2005heterogeneous}%
  \BibitemOpen
  \bibfield  {author} {\bibinfo {author} {\bibfnamefont {Malcolm~P}\
  \bibnamefont {Kennett}}, \bibinfo {author} {\bibfnamefont {Claudio}\
  \bibnamefont {Chamon}}, \ and\ \bibinfo {author} {\bibfnamefont {Leticia~F}\
  \bibnamefont {Cugliandolo}},\ }\bibfield  {title} {\enquote {\bibinfo {title}
  {Heterogeneous slow dynamics in a two dimensional doped classical
  antiferromagnet},}\ }\href@noop {} {\bibfield  {journal} {\bibinfo  {journal}
  {Phys. Rev. B}\ }\textbf {\bibinfo {volume} {72}},\ \bibinfo {pages} {024417}
  (\bibinfo {year} {2005})}\BibitemShut {NoStop}%
\bibitem [{\citenamefont {Rovere}\ \emph {et~al.}(1988)\citenamefont {Rovere},
  \citenamefont {Hermann},\ and\ \citenamefont {Binder}}]{rovere1988block}%
  \BibitemOpen
  \bibfield  {author} {\bibinfo {author} {\bibfnamefont {M}~\bibnamefont
  {Rovere}}, \bibinfo {author} {\bibfnamefont {DW}~\bibnamefont {Hermann}}, \
  and\ \bibinfo {author} {\bibfnamefont {K}~\bibnamefont {Binder}},\ }\bibfield
   {title} {\enquote {\bibinfo {title} {Block density distribution function
  analysis of two-dimensional lennard-jones fluids},}\ }\href@noop {}
  {\bibfield  {journal} {\bibinfo  {journal} {EPL (Europhysics Letters)}\
  }\textbf {\bibinfo {volume} {6}},\ \bibinfo {pages} {585} (\bibinfo {year}
  {1988})}\BibitemShut {NoStop}%
\bibitem [{\citenamefont {Block}\ \emph {et~al.}(2010)\citenamefont {Block},
  \citenamefont {Das}, \citenamefont {Oettel}, \citenamefont {Virnau},\ and\
  \citenamefont {Binder}}]{block2010curvature}%
  \BibitemOpen
  \bibfield  {author} {\bibinfo {author} {\bibfnamefont {Benjamin~J}\
  \bibnamefont {Block}}, \bibinfo {author} {\bibfnamefont {Subir~K}\
  \bibnamefont {Das}}, \bibinfo {author} {\bibfnamefont {Martin}\ \bibnamefont
  {Oettel}}, \bibinfo {author} {\bibfnamefont {Peter}\ \bibnamefont {Virnau}},
  \ and\ \bibinfo {author} {\bibfnamefont {Kurt}\ \bibnamefont {Binder}},\
  }\bibfield  {title} {\enquote {\bibinfo {title} {Curvature dependence of
  surface free energy of liquid drops and bubbles: A simulation study},}\
  }\href@noop {} {\bibfield  {journal} {\bibinfo  {journal} {J. Chem. Phys.}\
  }\textbf {\bibinfo {volume} {133}},\ \bibinfo {pages} {154702} (\bibinfo
  {year} {2010})}\BibitemShut {NoStop}%
\bibitem [{\citenamefont {Prestipino}\ \emph {et~al.}(2015)\citenamefont
  {Prestipino}, \citenamefont {Caccamo}, \citenamefont {Costa}, \citenamefont
  {Malescio},\ and\ \citenamefont {Muna{\`o}}}]{prestipino2015shapes}%
  \BibitemOpen
  \bibfield  {author} {\bibinfo {author} {\bibfnamefont {Santi}\ \bibnamefont
  {Prestipino}}, \bibinfo {author} {\bibfnamefont {Carlo}\ \bibnamefont
  {Caccamo}}, \bibinfo {author} {\bibfnamefont {Dino}\ \bibnamefont {Costa}},
  \bibinfo {author} {\bibfnamefont {Gianpietro}\ \bibnamefont {Malescio}}, \
  and\ \bibinfo {author} {\bibfnamefont {Gianmarco}\ \bibnamefont
  {Muna{\`o}}},\ }\bibfield  {title} {\enquote {\bibinfo {title} {Shapes of a
  liquid droplet in a periodic box},}\ }\href@noop {} {\bibfield  {journal}
  {\bibinfo  {journal} {Phys. Rev. E}\ }\textbf {\bibinfo {volume} {92}},\
  \bibinfo {pages} {022141} (\bibinfo {year} {2015})}\BibitemShut {NoStop}%
\bibitem [{\citenamefont {Vasisht}\ \emph {et~al.}(2011)\citenamefont
  {Vasisht}, \citenamefont {Saw},\ and\ \citenamefont {Sastry}}]{sastry_ll}%
  \BibitemOpen
  \bibfield  {author} {\bibinfo {author} {\bibfnamefont {Vishwas~V}\
  \bibnamefont {Vasisht}}, \bibinfo {author} {\bibfnamefont {Shibu}\
  \bibnamefont {Saw}}, \ and\ \bibinfo {author} {\bibfnamefont {Srikanth}\
  \bibnamefont {Sastry}},\ }\bibfield  {title} {\enquote {\bibinfo {title}
  {Liquid-liquid critical point in supercooled silicon},}\ }\href@noop {}
  {\bibfield  {journal} {\bibinfo  {journal} {Nature Phys.}\ }\textbf {\bibinfo
  {volume} {7}},\ \bibinfo {pages} {549--553} (\bibinfo {year}
  {2011})}\BibitemShut {NoStop}%
\bibitem [{\citenamefont {Stokely}\ \emph {et~al.}(2010)\citenamefont
  {Stokely}, \citenamefont {Mazza}, \citenamefont {Stanley},\ and\
  \citenamefont {Franzese}}]{stokely2010effect}%
  \BibitemOpen
  \bibfield  {author} {\bibinfo {author} {\bibfnamefont {Kevin}\ \bibnamefont
  {Stokely}}, \bibinfo {author} {\bibfnamefont {Marco~G}\ \bibnamefont
  {Mazza}}, \bibinfo {author} {\bibfnamefont {H~Eugene}\ \bibnamefont
  {Stanley}}, \ and\ \bibinfo {author} {\bibfnamefont {Giancarlo}\ \bibnamefont
  {Franzese}},\ }\bibfield  {title} {\enquote {\bibinfo {title} {Effect of
  hydrogen bond cooperativity on the behavior of water},}\ }\href@noop {}
  {\bibfield  {journal} {\bibinfo  {journal} {Proc. Natl. Acad. Sci. USA}\
  }\textbf {\bibinfo {volume} {107}},\ \bibinfo {pages} {1301--1306} (\bibinfo
  {year} {2010})}\BibitemShut {NoStop}%
\bibitem [{\citenamefont {Rovigatti}\ \emph {et~al.}(2017)\citenamefont
  {Rovigatti}, \citenamefont {Bianco}, \citenamefont {Tavares},\ and\
  \citenamefont {Sciortino}}]{lorenzo_spinodal}%
  \BibitemOpen
  \bibfield  {author} {\bibinfo {author} {\bibfnamefont {Lorenzo}\ \bibnamefont
  {Rovigatti}}, \bibinfo {author} {\bibfnamefont {Valentino}\ \bibnamefont
  {Bianco}}, \bibinfo {author} {\bibfnamefont {Jos?~Maria}\ \bibnamefont
  {Tavares}}, \ and\ \bibinfo {author} {\bibfnamefont {Francesco}\ \bibnamefont
  {Sciortino}},\ }\bibfield  {title} {\enquote {\bibinfo {title} {Re-entrant
  limits of stability of the liquid phase and the speedy scenario in colloidal
  model systems},}\ }\href {\doibase 10.1063/1.4974830} {\bibfield  {journal}
  {\bibinfo  {journal} {J. Chem. Phys.}\ }\textbf {\bibinfo {volume} {146}},\
  \bibinfo {pages} {041103} (\bibinfo {year} {2017})}\BibitemShut {NoStop}%
\bibitem [{\citenamefont {Molinero}\ \emph {et~al.}(2006)\citenamefont
  {Molinero}, \citenamefont {Sastry},\ and\ \citenamefont
  {Angell}}]{molinero2006tuning}%
  \BibitemOpen
  \bibfield  {author} {\bibinfo {author} {\bibfnamefont {Valeria}\ \bibnamefont
  {Molinero}}, \bibinfo {author} {\bibfnamefont {Srikanth}\ \bibnamefont
  {Sastry}}, \ and\ \bibinfo {author} {\bibfnamefont {C~Austen}\ \bibnamefont
  {Angell}},\ }\bibfield  {title} {\enquote {\bibinfo {title} {Tuning of
  tetrahedrality in a silicon potential yields a series of monatomic
  (metal-like) glass formers of very high fragility},}\ }\href@noop {}
  {\bibfield  {journal} {\bibinfo  {journal} {Phys. Rev. Lett.}\ }\textbf
  {\bibinfo {volume} {97}},\ \bibinfo {pages} {075701} (\bibinfo {year}
  {2006})}\BibitemShut {NoStop}%
\bibitem [{\citenamefont {Filion}\ \emph {et~al.}(2009)\citenamefont {Filion},
  \citenamefont {Marechal}, \citenamefont {van Oorschot}, \citenamefont {Pelt},
  \citenamefont {Smallenburg},\ and\ \citenamefont {Dijkstra}}]{boxshape1}%
  \BibitemOpen
  \bibfield  {author} {\bibinfo {author} {\bibfnamefont {Laura}\ \bibnamefont
  {Filion}}, \bibinfo {author} {\bibfnamefont {Matthieu}\ \bibnamefont
  {Marechal}}, \bibinfo {author} {\bibfnamefont {Bas}\ \bibnamefont {van
  Oorschot}}, \bibinfo {author} {\bibfnamefont {Dani{\"e}l}\ \bibnamefont
  {Pelt}}, \bibinfo {author} {\bibfnamefont {Frank}\ \bibnamefont
  {Smallenburg}}, \ and\ \bibinfo {author} {\bibfnamefont {Marjolein}\
  \bibnamefont {Dijkstra}},\ }\bibfield  {title} {\enquote {\bibinfo {title}
  {Efficient method for predicting crystal structures at finite temperature:
  Variable box shape simulations},}\ }\href@noop {} {\bibfield  {journal}
  {\bibinfo  {journal} {Phys. Rev. Lett.}\ }\textbf {\bibinfo {volume} {103}},\
  \bibinfo {pages} {188302} (\bibinfo {year} {2009})}\BibitemShut {NoStop}%
\bibitem [{\citenamefont {de~Graaf}\ \emph {et~al.}(2012)\citenamefont
  {de~Graaf}, \citenamefont {Filion}, \citenamefont {Marechal}, \citenamefont
  {van Roij},\ and\ \citenamefont {Dijkstra}}]{boxshape2}%
  \BibitemOpen
  \bibfield  {author} {\bibinfo {author} {\bibfnamefont {Joost}\ \bibnamefont
  {de~Graaf}}, \bibinfo {author} {\bibfnamefont {Laura}\ \bibnamefont
  {Filion}}, \bibinfo {author} {\bibfnamefont {Matthieu}\ \bibnamefont
  {Marechal}}, \bibinfo {author} {\bibfnamefont {Ren{\'e}}\ \bibnamefont {van
  Roij}}, \ and\ \bibinfo {author} {\bibfnamefont {Marjolein}\ \bibnamefont
  {Dijkstra}},\ }\bibfield  {title} {\enquote {\bibinfo {title}
  {Crystal-structure prediction via the floppy-box monte carlo algorithm:
  Method and application to hard (non) convex particles},}\ }\href@noop {}
  {\bibfield  {journal} {\bibinfo  {journal} {J. Chem. Phys.}\ }\textbf
  {\bibinfo {volume} {137}},\ \bibinfo {pages} {214101} (\bibinfo {year}
  {2012})}\BibitemShut {NoStop}%
\bibitem [{\citenamefont {Kofke}(1993)}]{kofke}%
  \BibitemOpen
  \bibfield  {author} {\bibinfo {author} {\bibfnamefont {David~A}\ \bibnamefont
  {Kofke}},\ }\bibfield  {title} {\enquote {\bibinfo {title} {Direct evaluation
  of phase coexistence by molecular simulation via integration along the
  saturation line},}\ }\href@noop {} {\bibfield  {journal} {\bibinfo  {journal}
  {J. Chem. Phys.}\ }\textbf {\bibinfo {volume} {98}},\ \bibinfo {pages}
  {4149--4162} (\bibinfo {year} {1993})}\BibitemShut {NoStop}%
\bibitem [{\citenamefont {Vega}\ \emph {et~al.}(2008)\citenamefont {Vega},
  \citenamefont {Sanz}, \citenamefont {Abascal},\ and\ \citenamefont
  {Noya}}]{Vega}%
  \BibitemOpen
  \bibfield  {author} {\bibinfo {author} {\bibfnamefont {C}~\bibnamefont
  {Vega}}, \bibinfo {author} {\bibfnamefont {E}~\bibnamefont {Sanz}}, \bibinfo
  {author} {\bibfnamefont {JLF}\ \bibnamefont {Abascal}}, \ and\ \bibinfo
  {author} {\bibfnamefont {EG}~\bibnamefont {Noya}},\ }\bibfield  {title}
  {\enquote {\bibinfo {title} {Determination of phase diagrams via computer
  simulation: methodology and applications to water, electrolytes and
  proteins},}\ }\href@noop {} {\bibfield  {journal} {\bibinfo  {journal} {J.
  Phys.: Condens. Matter}\ }\textbf {\bibinfo {volume} {20}},\ \bibinfo {pages}
  {153101} (\bibinfo {year} {2008})}\BibitemShut {NoStop}%
\bibitem [{\citenamefont {Panagiotopoulos}(2000)}]{reweight}%
  \BibitemOpen
  \bibfield  {author} {\bibinfo {author} {\bibfnamefont {Athanassios~Z}\
  \bibnamefont {Panagiotopoulos}},\ }\bibfield  {title} {\enquote {\bibinfo
  {title} {Monte carlo methods for phase equilibria of fluids},}\ }\href@noop
  {} {\bibfield  {journal} {\bibinfo  {journal} {J. Phys.: Condens. Matter}\
  }\textbf {\bibinfo {volume} {12}},\ \bibinfo {pages} {R25} (\bibinfo {year}
  {2000})}\BibitemShut {NoStop}%
\bibitem [{\citenamefont {Tsypin}\ and\ \citenamefont
  {Bl{\"o}te}(2000)}]{tsypin}%
  \BibitemOpen
  \bibfield  {author} {\bibinfo {author} {\bibfnamefont {MM}~\bibnamefont
  {Tsypin}}\ and\ \bibinfo {author} {\bibfnamefont {HWJ}\ \bibnamefont
  {Bl{\"o}te}},\ }\bibfield  {title} {\enquote {\bibinfo {title} {Probability
  distribution of the order parameter for the three-dimensional ising-model
  universality class: A high-precision monte carlo study},}\ }\href@noop {}
  {\bibfield  {journal} {\bibinfo  {journal} {Phys. Rev. E}\ }\textbf {\bibinfo
  {volume} {62}},\ \bibinfo {pages} {73} (\bibinfo {year} {2000})}\BibitemShut
  {NoStop}%
\bibitem [{\citenamefont {Romano}\ \emph {et~al.}(2011)\citenamefont {Romano},
  \citenamefont {Sanz},\ and\ \citenamefont {Sciortino}}]{romano_tetrahedral}%
  \BibitemOpen
  \bibfield  {author} {\bibinfo {author} {\bibfnamefont {Flavio}\ \bibnamefont
  {Romano}}, \bibinfo {author} {\bibfnamefont {Eduardo}\ \bibnamefont {Sanz}},
  \ and\ \bibinfo {author} {\bibfnamefont {Francesco}\ \bibnamefont
  {Sciortino}},\ }\bibfield  {title} {\enquote {\bibinfo {title}
  {Crystallization of tetrahedral patchy particles in silico},}\ }\href
  {\doibase DOI:10.1063/1.3578182} {\bibfield  {journal} {\bibinfo  {journal}
  {J. Chem. Phys.}\ }\textbf {\bibinfo {volume} {134}},\ \bibinfo {pages}
  {174502} (\bibinfo {year} {2011})}\BibitemShut {NoStop}%
\end{thebibliography}

\beginsupplement

\section*{Supplementary Information}

\subsection*{Three dimensional $T$-$P$-$\lambda$ phase diagram}

\begin{figure*}[!b]
\centering
 \includegraphics[width=13cm,clip]{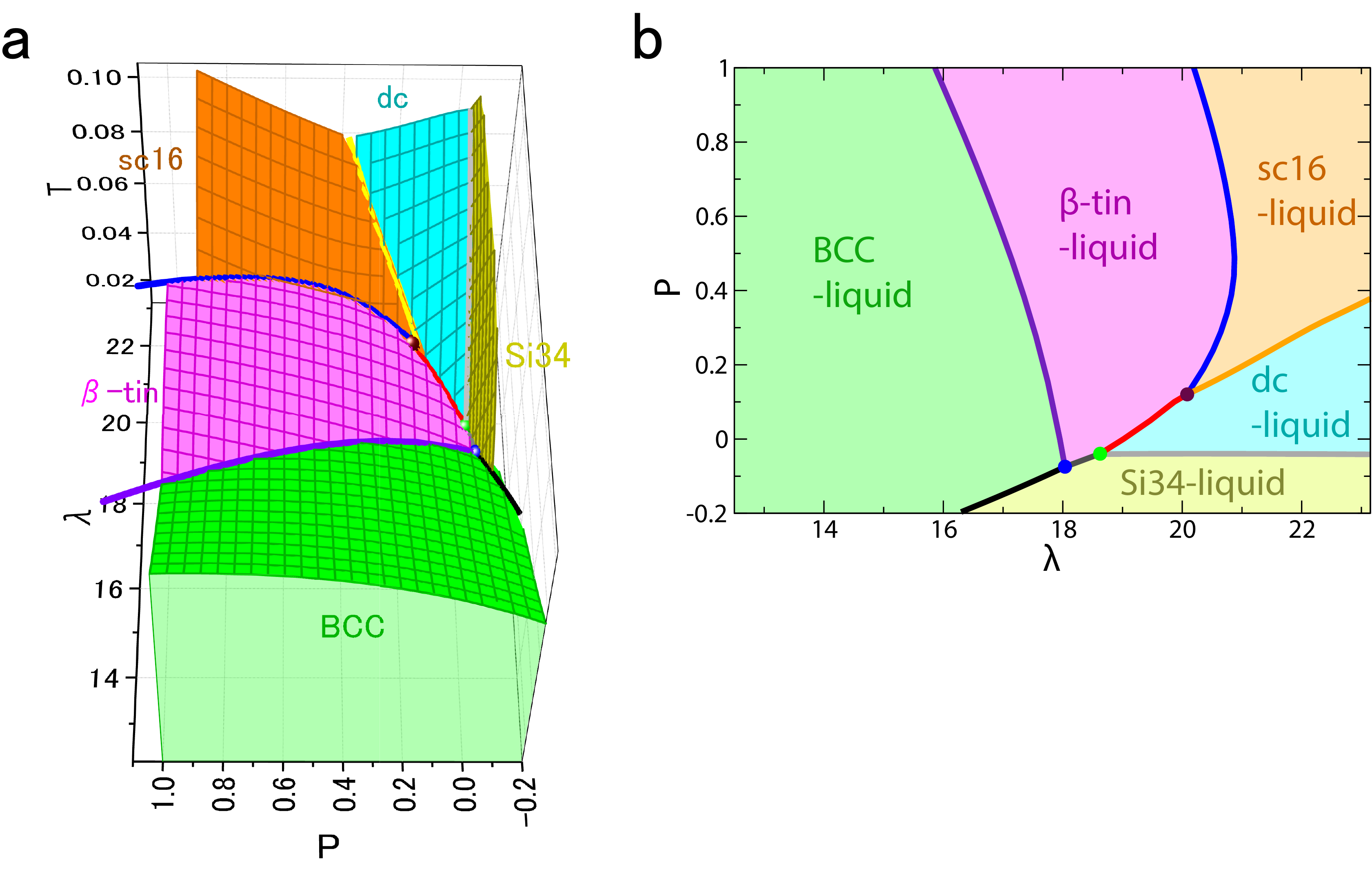}
 \caption{Phase diagram of a system interacting with the SW potential including negative pressures. ($A$) The $\lambda$-$P$-$T$ phase diagram. The green, pink, turquoise, orange and yellow surfaces are liquid-BCC, liquid-$\beta$-tin, liquid-dc, liquid-sc16 and liquid-Si34 coexisting surfaces. The purple, red, yellow, blue, grey, black and dark grey lines are liquid-BCC-$\beta$-tin, liquid-$\beta$-tin-dc, liquid-dc-sc16, liquid-$\beta$-tin-sc16, liquid-dc-Si34, liquid-BCC-Si34, liquid-$\beta$-tin-Si34 coexisting lines. The brown, green and blue points are liquid-$\beta$-tin-dc-sc16, liquid-$\beta$-tin-dc-Si34 and liquid-BCC-$\beta$-tin-Si34 coexisting points. ($B$) The projection of the coexisting regions onto the $\lambda$-$P$ plane. The green, pink, turquoise, orange, yellow regions are the projection of BCC-liquid, $\beta$-tin-liquid, dc-liquid, sc16-liquid, Si34-liquid surfaces into  $\lambda$-$P$ plane respectively. 
}
 \label{fig:3dpd}
\end{figure*}%

A three dimensional extension of the phase diagram of the SW potential can be obtained by promoting $\lambda$ as an effective thermodynamic
parameter, acting like an external field. In Ref.~\cite{akahane_quadruple} we computed the full phase diagram of the three-dimensional SW model
in which $P$, $T$ and $\lambda$ are the parameters. The following crystalline structures were found to be stable:
bcc at low values of $\lambda$, $\beta$-tin at intermediate values of $\lambda$, $dc$ at high values of $\lambda$ and small $P$,
and sc16 at high values of $\lambda$ and high $P$. This last phase, the sc16 crystal, was first discovered in Ref.~\cite{romano2014novel}, and opened
a new intriguing scenario for the phase behavior of the SW potential, namely the possibility of a quadruple point, where dc, $\beta$-tin, sc16, and the
fluid phase would all coexist at the same thermodynamic conditions. This was indeed found in Ref.~\cite{akahane_quadruple}. In the $T$-$P$ plane, a SW model
with $\lambda\sim 20.08$ consequently shows a quadruple point. We note that this value is close to that of germanium ($\lambda=20.0$).

Here we extend the phase diagram to negative pressures. Negative pressures are of great interest for at least two important reasons: 1) they stabilize clathrate lattices, which are crystalline structures with voids that can accommodate guest molecules, and are studied for energy storage, carbon dioxide sequestration, separation and natural gas storage~\cite{florusse2004stable,lee2005tuning,chatti2005benefits,struzhkin2007hydrogen}; 2) contrasting theories of the thermodynamic anomalies (in particular for the case of water) can be tested in the negative pressure region, where they make different predictions.
We thus include several clathrate structures in our thermodynamic calculations: the structures are the clathrates Si34, Si46 and Si136~\cite{romano2014novel}.
In Ref.~\cite{romano2014novel} it was shown that the stable crystal at negative pressure is the Si34 clathrate for the SW model parameterizations of silicon and water.
We use data in Ref.~\cite{romano2014novel} as the starting point and extend Si34-liquid coexistence lines to lower $\lambda$ and lower $P$ as well as dc-liquid, BCC-liquid, $\beta$-tin-liquid coexisting lines. 

In Fig.~\ref{fig:3dpd}$A$, each surface represents a coexistence surface between the liquid and the corresponding crystal.
Thick lines are triple lines, where two crystalline phases and the liquid phase coexist.
To aid the visualization, we also plot in Fig.~\ref{fig:3dpd}$B$ a projection of the coexistence surfaces onto the $(P,\lambda)$ plane.
Figure~\ref{fig:3dpd} shows that indeed the clathrate Si34 is the thermodynamic stable phase at negative pressures, where it can
coexist with the bcc crystal along a triple line.
At negative pressure, the BCC (body-centered cubic) phase is stable at lower $\lambda$ and the Si34 phase is stable at higher $\lambda$.
Interestingly two new quadruple points emerge at negative pressures: the first one is found at the coexistence between
the dc, $\beta$-tin, sc16, and liquid phases, and the second one at the coexistence between $\beta$-tin, Si34, liquid and BCC phases. Quadruple points in a one-component system are possible due to
the extension of the thermodynamic parameter space to include the $\lambda$ parameter, for which the Gibbs rule of phases has to be generalized as $F=C-N+R$, where $F$ are the degrees of freedom, $R$ is the number of independent intensive parameters, $C$ is the number of chemical components, $N$ is the number of phases: for a quadruple point ($N=4$), in a one component systems ($C=1$) with $T$, $P$, and $\lambda$ as intensive parameters ($R=3$), we have $F=0$, which denotes a dimensionless point ($T_\text{QP},P_\text{QP},\lambda_\text{QP}$). For a detailed study of the thermodynamic properties of quadruple points see Ref.~\cite{akahane_quadruple}.

The phase diagram of Fig.~\ref{fig:3dpd} cannot be extended to lower pressure due to the instability of the liquid phase against vapor nucleation.

\subsection*{$\lambda$-dependence of liquid-solid coexistence}

\begin{figure*}[!ht]
\centering
 \includegraphics[width=11.8cm,clip]{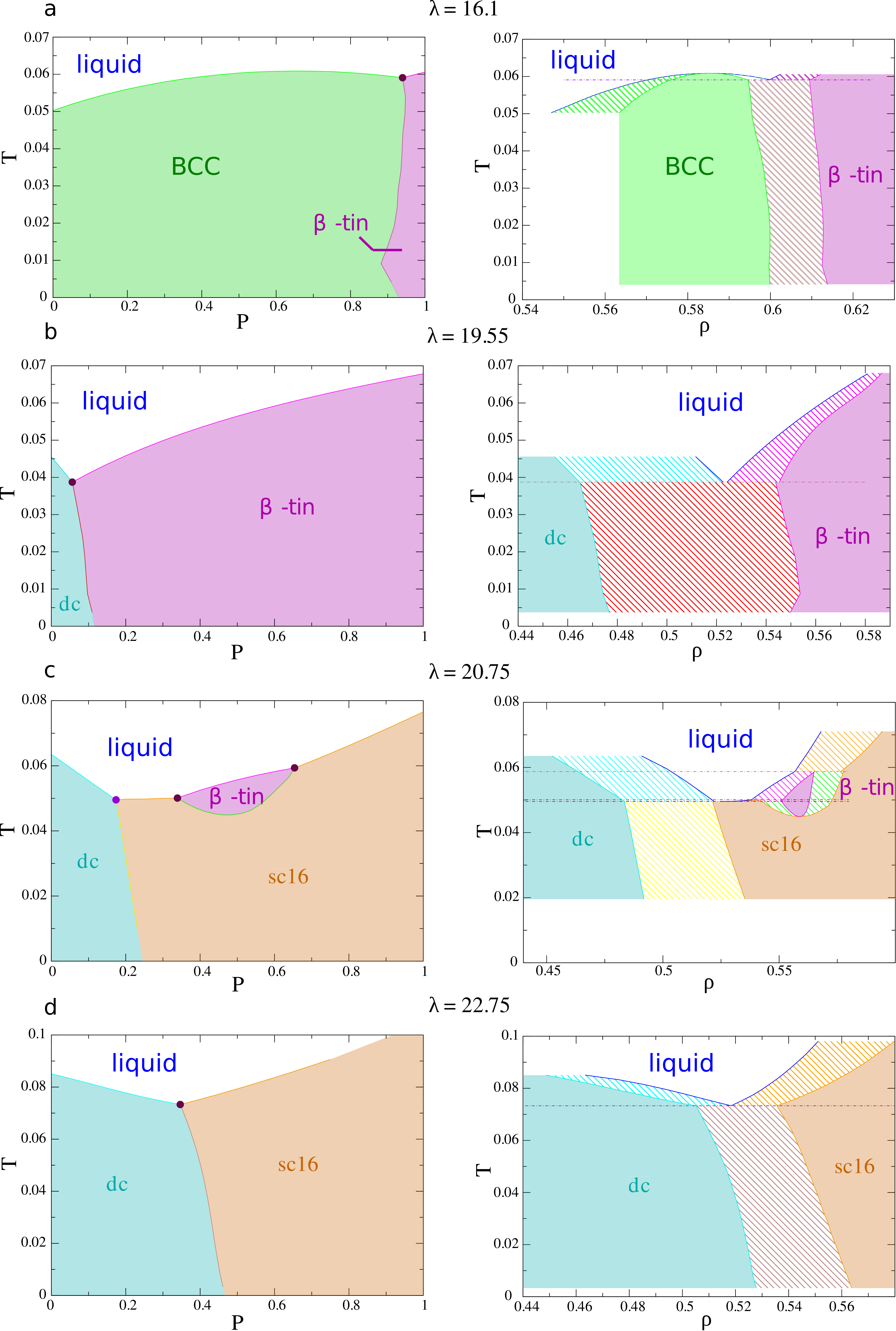}
 \caption{{\bf $P$-$T$ and $\rho$-$T$ phase diagrams of the SW potential at $\lambda=16.1, 19.55, 20.75, 22.75$.} The bcc, $\beta$-tin, dc, and sc16 phases are stable in green, pink, blue, and orange regions respectively. In $P$-$T$ phase diagrams, circle points are triple points. In $\rho$-$T$ phase diagrams, regions with diagonal lines are the coexisting regions between two phases. The horizontal lines denote the temperatures of the corresponding triple points.}
 \label{fig:pd}
\end{figure*}%

In this Section we focus on models with fixed $\lambda$ and study the different types of phase diagrams that
characterize the generalized SW potential. By varying $\lambda$, the main changes to the phase diagram are depicted in
Fig.~\ref{fig:pd}, where both the $P$-$T$ and $\rho$-$T$ planes are computed for selected values of $\lambda$.
The choices correspond to the following phase diagram types: in order of increasing $P$,
BCC-$\beta$-tin ($\lambda=16.1$, panel a), dc-$\beta$-tin ($\lambda=19.55$, panel b), dc-sc16-$\beta$-tin-sc16 ($\lambda=20.75$, panel c), dc-sc16 ($\lambda=22.75$, panel d).
We can observe the following trend: by decreasing $\lambda$, a re-entrant $\beta$-tin phase appears inside the sc16 stability region. Between $\lambda=20.75$
and $\lambda=19.55$ the $\beta$-tin phase expands inside the sc16 region and pushing the diamond phase to lower pressures. At $\lambda=20.08$ a quadruple
point appears where dc, sc16 and $\beta$-tin coexist with the fluid phase~\cite{akahane_quadruple}. For $\lambda<20.08$ the sc16 becomes metastable at intermediate pressures. Further decreasing $\lambda$ consolidates the stability of the $\beta$-tin phase, which eventually overcomes the dc phase at $P=0$~\cite{molinero2006tuning,romano2014novel}.
Finally, for $\lambda<18$ the BCC phase starts emerging at lower pressures, eventually becoming the dominant phase at $\lambda=16.1$.

For every $\lambda$, the density of the different crystals shows the following trend:
the dc phase has the lowest $\rho$, followed by BCC whose density is always lower than the one of the $\beta$-tin crystal. The $\beta$-tin and sc16 are the
high density phases, but the sc16 phase is stable over a wider range of $\rho$, which is the reason why it eventually preempts the 
$\beta$-tin
phase at higher values of $\lambda$.

The phase diagrams in Fig.~\ref{fig:pd} show the richness in physical behavior of the SW potential, from open crystalline structures,
to re-entrant solid-solid transitions. They also show that the quadruple point that was introduced in Ref.~\cite{akahane_quadruple} comes
from the merging of three triple points: the $\beta$-tin-sc16-liquid and dc-sc16-liquid triple points at $\lambda>20.08$ with the
dc-$\beta$-tin-liquid triple point at $\lambda<20.08$.

\subsection*{Liquid-gas critical points}

\begin{figure*}[!t]
\centering
 \includegraphics[width=13cm,clip]{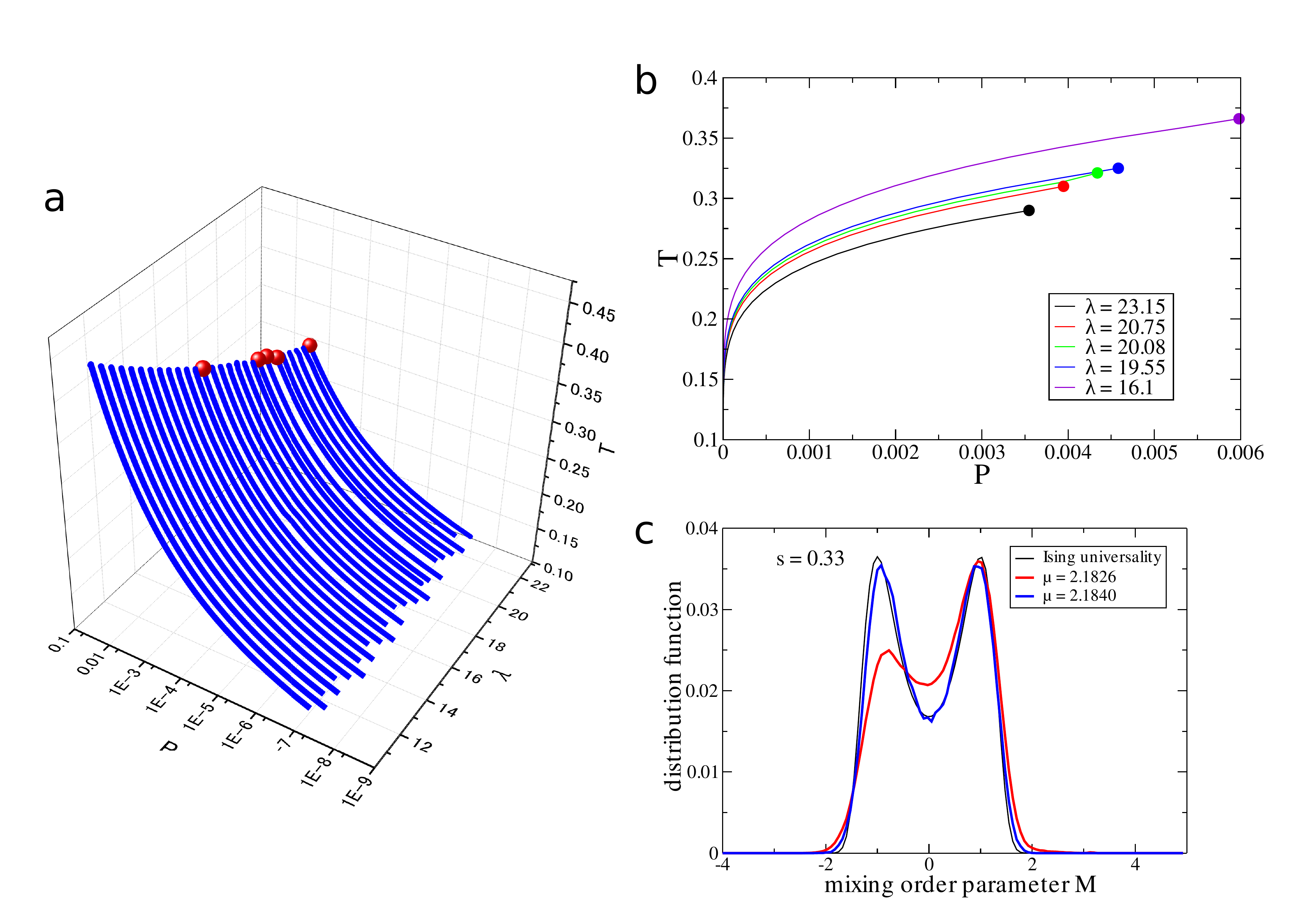}
 \caption{The liquid-gas critical points and coexisting lines of a system with the SW potential. 
($A$) In $\lambda$-$P$-$T$ space. Pressures are in logarithmic scale. The red points are critical points. The blue lines are coexisting lines. ($B$) In $P$-$T$ plane.
The black, red, green, blue, and purple lines and points are coexisting lines and critical points at $\lambda=23.15, 20.75, 20.08, 19.55,$ and $16.1$ respectively.
($C$) An example of the distribution functions of the mixing order parameter $M$. The mixing parameter is $m=0.33$.
The blue curve is the one at $\lambda=20.75$, $T=0.310$, and $\mu=2.1840$ which is re-weighted from the red one at $\mu=2.1826$. The black curve is the universal distribution of the mixing order parameter for the Ising universality class.}
 \label{fig:lg}
\end{figure*}

As a first step, we consider here the liquid-gas coexistence line, and the critical point. In the ($T$,$P$,$\lambda$) space, the generalized SW
model will have a line of critical points and a liquid-gas coexistence surface. To locate critical points we conduct grand canonical ensemble simulations,
where the number of particles fluctuates around the equilibrium value set by the chemical potential, $\mu$.
By exploring the $\mu$-$T$ phase space we compute the $\rho$ and internal energy ($u$) histograms. The critical point
is located by finding the exact thermodynamic conditions at which the mixing order parameter $M=\rho+m\,u$ (with $m$ the mixing
parameter) coincides with the universal three-dimensional Ising universality class distribution~\cite{tsypin}.
In order to explore the probability distributions for small displacements of the thermodynamic conditions we employ
histogram re-weighting~\cite{reweight}. The steps are the following.
First, we obtain the histograms of densities and internal energies in grand canonical ensembles ${\mu}VT$.
In the next step, we re-weight those histograms by multiplying $e^{({\mu}'-{\mu}){\beta}N}$ and obtain histograms of densities and internal energies in ensemble ${\mu}'VT$.
In the third step, we calculate the distribution functions $P(M)$ of order parameter $M$ with changing the mixing parameter $m$.
Finally, we calculate mean squared errors between $P(M)$ and standardized Ising universality curve~\cite{tsypin} and
choose the set of $(T,\mu,m)$ which minimize the error. We show an example of this fitting procedure in Fig.~\ref{fig:lg}$C$ for $\lambda=20.75$ and $T=0.310$: first
we obtain the distribution function $P(M)$ for $\mu=2.1826$ (red curve), then, by histogram re-weighting, we find the values of $\mu=2.1840$ and $m=0.33$ (blue curve) as the best fit to the Ising universality curve (black curve).
Following the above mentioned procedure we compute the critical point for different values of $\lambda$.

The critical points are plotted as full symbols
in Fig.~\ref{fig:lg}$A$ and $B$. We see that increasing $\lambda$ shifts the critical point to both lower temperatures and pressures.
We can compare these results to the ones obtained with tetrahedral Patchy Particles, which are colloidal particles with directional interactions~\cite{romano_tetrahedral}. In the case of Patchy
Particles, the controlling parameter is the angular width of the patches, $\phi$, that controls the bonding volume of the interaction and deviations from tetrahedrality, in a way that resembles our $\lambda$ parameter: decreasing $\phi$ and increasing $\lambda$ both produce a stronger tetrahedral local arrangement in Patchy Particles and the SW model respectively. In Patchy Particles, decreasing $\phi$ suppresses the critical point until it becomes metastable to crystallization. In the SW model, increasing $\lambda$ also suppresses the critical temperature, but here it remains always stable: comparing the temperature range of the critical points in Fig.~\ref{fig:lg}$B$, with the solid coexistence points of Fig.~\ref{fig:pd}, we see that the critical point always remains far above the melting lines. The metastability of the critical point in Patchy Particles is due to the potential being considerably shorter ranged than the SW potential. Regarding the critical pressure, while the critical pressure 
increases with decreasing $\phi$, for the SW model, an increase of tetrahedrality also reduces the critical pressure.

Figures~\ref{fig:lg}$A$ and $B$ also report the liquid-gas coexistence points for several value of $\lambda$. These lines are obtained with
the Gibbs-Duhem integration. First, successive-umbrella simulations are conducted in proximity of the critical point to determine
the coexistence point directly from the distribution function of the order parameter (i.e. coexistence is defined when the area under
the gas and liquid peak of $P(\rho)$ are equal). Then, starting from this coexistence point, the Gibbs-Duhem integration is used to
compute the next coexistence point at lower pressures, repeating the process iteratively. The advantage of using successive umbrella
simulations is that it gives the coexistence point for temperatures in which the liquid-gas free energy barrier is too low for
Gibbs-Duhem integration (which requires long metastability of both the liquid and gas phases).

\subsection*{Equation of state near retracing spinodal}

\begin{figure*}
 \centering
 \includegraphics[width=13cm,clip]{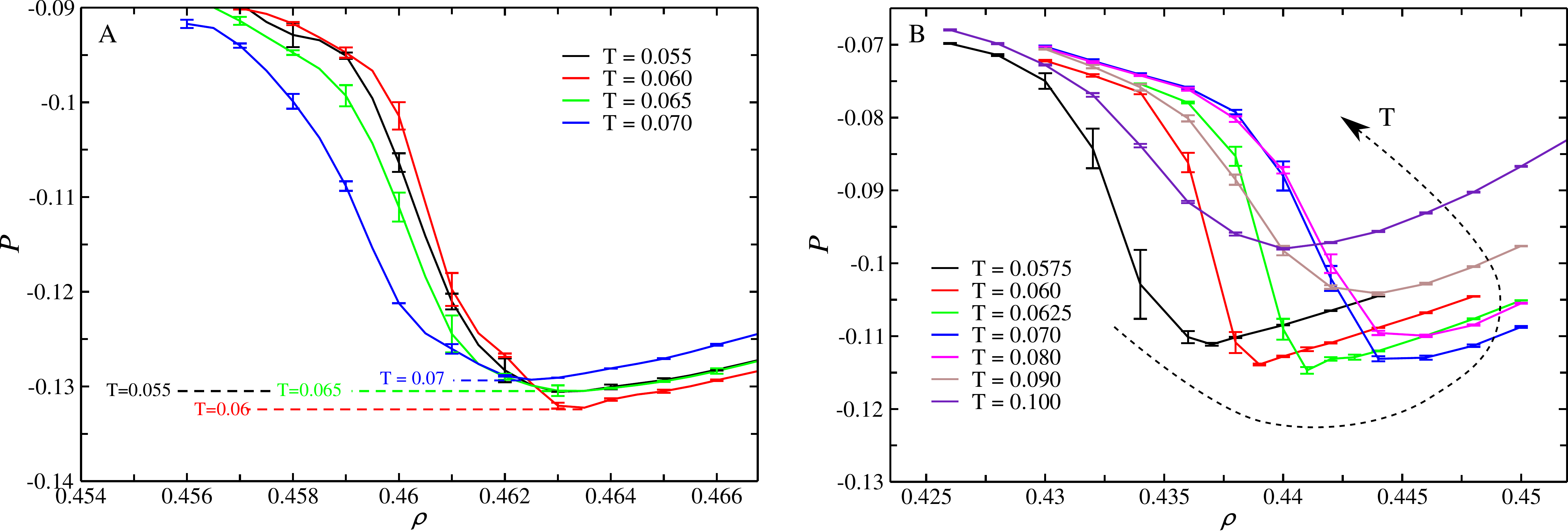}
 \caption{Equation of state for $\lambda=19.55$ (A, left) and $\lambda=20.75$ (B, right) for different $T$ around the re-tracing spinodal. Each state point corresponds to the average over $5$ independent trajectories, while checking for the absence of crystal nucleation.}
 \label{fig:eos}
\end{figure*}

The retracing spinodals reported in the insets of Fig.~5 of the main text are obtained by computing the equation of state $P(\rho)$ for different $T$ in the NVT ensemble. In Fig.~\ref{fig:eos} we report the $P(\rho)$ curves obtained from averaging over $5$ independent trajectories. The spinodal line is obtained from the conditions $\frac{d\,P}{d\,\rho}=0$ and $\frac{d^2\,P}{d\,\rho^2}>0$, where derivatives are computed from the a cubic spline interpolation of the points in Fig.~\ref{fig:eos}.

\end{document}